\documentclass[final,3p,times,twocolumn]{elsarticle}

\usepackage{amssymb}
\usepackage{amsmath}
\usepackage{booktabs}
\usepackage{lineno}

\usepackage[dvipsnames]{xcolor}

\journal{Astroparticle Physics}

\begin{document}

\begin{frontmatter}



\title{Likelihood-based reconstruction of muon lateral distribution function using combined integrator and binary detector modes}



\author[iteda]{A.D. Supanitsky} 
\author[iteda]{D. Ravignani}
\author[iteda]{V.V. Kizakke Covilakam}

\affiliation[iteda]{organization={Instituto de Tecnologías en Detección y Astropartículas (ITeDA, CNEA-CONICET-UNSAM)},
            addressline={Av. Gral Paz 1499}, 
            city={Gral. San Martín},
            postcode={B1650KNA}, 
            state={Buenos Aires},
            country={Argentina}}

\begin{abstract}

The origin of ultra-high-energy cosmic rays, with energies $E \geq 10^{18}$ eV, remains unknown. Among 
the key observables used to investigate their nature are the energy spectrum, the arrival direction 
distribution, and the composition as a function of energy. The composition of the primary cosmic ray is 
inferred from properties of the extensive air showers they initiate, particularly from parameters sensitive 
to the primary mass. The most sensitive parameters to the primary mass are the atmospheric depth of the 
shower maximum, typically measured with fluorescence telescopes, and the muon content of the shower, measured 
using dedicated muon detectors. A commonly used observable in composition studies is the muon density at a 
fixed distance from the shower axis, derived by evaluating the reconstructed muon lateral distribution 
function (MLDF) at a reference distance. A specific type of muon detector features two acquisition modes: 
binary and integrator (commonly referred to as ADC mode, for Analog-to-Digital Converter). The binary mode 
allows for direct muon counting, while the ADC mode infers the muon number from the integrated signal of the
detector response. Existing methods reconstruct the MLDF using data from either acquisition mode individually, 
or by combining both, but usually assigning a single mode per detector station in a given event. This work
presents a novel method to reconstruct the MLDF based on a likelihood approach that simultaneously incorporates 
data from both acquisition modes at each detector station. We apply our method to the underground muon detectors 
of the Pierre Auger Observatory as a case study. However, this general approach can be applied to future 
detectors with dual acquisition capabilities. Our results demonstrate that the combined method outperforms 
traditional techniques that rely solely on either binary or ADC mode data.

\end{abstract}




\end{frontmatter}



\section{Introduction}
\label{Intro}

The origin of ultra-high-energy cosmic rays (UHECRs, $E \geq 10^{18}$ eV) remains one of the major open questions 
in astroparticle physics. Nonetheless, significant progress has been made in recent years, largely due to data 
collected by current observatories: the Pierre Auger Observatory (hereafter referred to as Auger) \cite{Auger:15} 
in the Southern Hemisphere and the Telescope Array (TA) \cite{TA:08} in the Northern Hemisphere. Given their extremely 
low flux, UHECRs cannot be detected directly. Instead, they are observed indirectly through the extensive air showers 
they produce upon interacting with atmospheric molecules. As a result, key properties such as the primary energy, nuclear 
composition, and arrival direction of the cosmic ray must be reconstructed from the characteristics of the resulting air 
shower.

The three primary observables used to investigate the nature of UHECRs are the energy spectrum, the composition profile, 
and the distribution of their arrival directions. Since the composition cannot be measured directly, it must be inferred 
from air shower observables. Several parameters are sensitive to the mass of the primary particle, with the most prominent 
being the atmospheric depth at which the shower reaches its maximum development ($X_{\textrm{max}}$) and the muon content 
of the shower \cite{Supa:22}. Typically, the muon density at a fixed distance from the shower axis is employed as a 
composition-sensitive parameter. While $X_{\textrm{max}}$ is measured using fluorescence telescopes, the muon content is 
determined using dedicated muon detectors.

Composition information is crucial for understanding the origin of cosmic rays. In particular, it is believed that the 
composition plays a key role in studying the transition from galactic to extragalactic cosmic rays \cite{Supa:22, Aloisio:12}, 
as well as in determining the nature of the flux suppression observed around $10^{19.7}$ eV \cite{Kampert:13, Auger:16}. 
Determining the composition of UHECRs is a challenging task, as it relies on comparisons between air shower simulations and 
observational data. Since hadronic interactions at very high energies are not well known, these simulations depend on models 
that extrapolate results from accelerator experiments to much higher energies. This approach introduces significant systematic 
uncertainties into the composition analysis even when models updated with Large Hadron Collider (LHC) data are considered. Moreover, 
several experiments have reported a discrepancy between simulated air showers and observational data \cite{Whisp:23}. This 
discrepancy is commonly interpreted as a muon deficit in air showers simulated using post-LHC high-energy hadronic interaction 
models. Although this muon deficit constitutes a significant source of systematic uncertainty in composition analyses based 
on the muon content of the showers, it is expected that future generations of high-energy hadronic interaction models will help 
to reduce or resolve this issue.

The muon content of extensive air showers can be directly measured using dedicated muon detectors, which are shielded to suppress 
contamination from the electromagnetic component of the showers. A variety of such detectors have been developed and some of 
them  are currently in use to quantify the muon component \cite{Whisp:19}. Notable examples include those employed by Auger 
(currently operational) \cite{AMIGA:21} and the Akeno Giant Air Shower Array (AGASA), which is no longer in operation \cite{Agasa:95}. 
These detectors are segmented and operate in two acquisition modes: binary mode and ADC (analog-to-digital converter) mode. In binary 
mode, a muon is counted when its signal exceeds a predefined amplitude threshold within a specified time window in any segment. A 
significant limitation of this mode is known as the pile-up effect \cite{Supa:08}. When multiple muons hit the same segment during 
the inhibition window, they are counted as a single event. In contrast, the ADC (or integrator) mode measures the total integrated 
signal deposited by muons across all segments. The total muon count is then estimated by dividing this integrated signal by the 
expected signal from a single muon. Both acquisition modes are susceptible to saturation. In binary mode, saturation occurs when all 
segments are triggered simultaneously within a time window, leading to an undercounting of muons \cite{Gesualdi:22}. In ADC mode, 
saturation happens when the pulse amplitude exceeds the dynamic range of the system.

Several methods have been developed to reconstruct the muon lateral distribution function (MLDF) for the Auger underground 
muon detectors. The approach described in Ref.~\cite{Supa:08} utilizes the binary mode. This method employs a Poisson likelihood, 
coupled with a correction for the pile-up effect, to infer the number of muons striking a given detector. When the uncertainty 
from the pile-up correction exceeds the Poisson uncertainty, which occurs when a large number of muons hit a detector within a 
time interval given by the time resolution of the electronics, a lower limit is applied. Ref.~\cite{Ravignani:15} introduced a new 
likelihood for the binary mode that avoids approximations for infinite inhibition windows. This method was then extended in 
Ref.~\cite{Ravignani:16} to incorporate finite inhibition windows, though in this case, the likelihood function is approximated. 
More recently, in Ref.~\cite{Varada:23} an MLDF reconstruction method based on the ADC mode has been introduced. This study has 
also introduced a novel hybrid method that strategically combines both acquisition modes. In this case, the ADC mode is used 
only for stations with high muon counts, while the binary mode is used for the remaining stations involved in the event.

In this work, we present a new method for reconstructing the MLDF that uses a likelihood function combining data from both acquisition 
modes, binary and ADC, for most of the stations involved in a given event. Although the method is suitable for use in current and 
future arrays of muon detectors, it is applied here to the array of underground muon detectors at Auger, which serves as a case study. 
We demonstrate that this combined approach outperforms methods based solely on either the binary mode or the ADC mode. It is worth 
mentioning that this new method can be relevant in a very different context like photon counting in lidar measurements \cite{Darko:11}.   
Section \ref{sec:ComLike} introduces the new likelihood function. In Section \ref{sec:EstMu}, we examine the estimation of the mean number 
of muons at the station level using the new likelihood. Section \ref{sec:Rec} presents the development of the combined reconstruction 
method and evaluates its performance, including a discussion on the reconstruction of saturated events. Finally, Section \ref{sec:Conc} 
summarizes our conclusions.

\section{The combined likelihood}
\label{sec:ComLike}

The MLDF is defined as the average number of muons at ground level produced in an air shower at a given distance from the shower 
axis. It can be expressed as $\mu = \rho_\mu \, A \cos\theta$, where $\rho_\mu$ is the muon density at the considered distance, 
$A$ is the detector area, and $\theta$ is the zenith angle of the shower. When a shower is detected, several muon detectors are
typically triggered. Each participating detector is struck by given a number of muons, denoted by $n$, thereby providing a discrete 
sampling of the MLDF at its distance from the shower axis. In the case of an ideal detector, $n$ could be measured directly. 
However, real muon detectors record observables that depend on $n$. For segmented muon detectors operating in binary or ADC 
acquisition modes, the observables are the number of activated bars (or segments), denoted by $k$, and the total collected 
charge, denoted by $Q$, respectively. In binary mode, a bar is considered activated when the time series of logical signals 
generated by the discriminator electronics matches a predefined pattern associated with a muon. In ADC mode, the electronics 
measure the total signal, i.e., the sum of signals from all segments, as a function of time. This signal is generally recorded 
in ADC counts, and the total charge $Q$ is obtained by summing the counts across all time bins, which is the approach adopted 
in this work. 

Regarding the binary mode the distribution function of $k$ for the case of an infinite inhibition window is given by 
\cite{Ravignani:15}
\begin{equation}
P_\textrm{{\scriptsize B}} (k;\mu) = \binom{n_{\textrm{s}}}{k} \exp(-\mu) \, \left(\exp(\mu/n_{\textrm{s}})-1 \right)^k,
\label{fB}
\end{equation}
where $n_{\textrm{s}}$ is the total number of bars of the detector.

The likelihood function corresponding to the ADC channel has been presented in Ref.~\cite{Varada:23}, a brief review of its derivation 
is provided below to ensure a complete and self-contained presentation of the combined likelihood function. The charge deposited by a 
single muon is described by a log-normal distribution \cite{Botti:19}, given by $f(q) = \textrm{LN}(q;m,\theta^2)$. Typical values for 
the parameters corresponding to the low gain channel of the ADC of the Auger underground muon detectors are $\exp(m) = \exp(5)$ ADC 
counts and $\theta = 0.5$ \cite{Botti:19}. Due to the linear response of the ADC \cite{Eng:20}, the total charge for the case in which 
$n$ muons hit a given muon detector is given by $Q=\sum_{i=1}^n q_i$, where $q_i$ are $n$ independent random samples of $f(q)$. 

Since $\theta^2 < 1$, the distribution function of $Q$ can also be approximated by a log-normal distribution 
\cite{Varada:23}
\begin{equation}
\label{LNn}
g_{\textrm{{\scriptsize ADC}}}(Q;n) = \frac{1}{\sqrt{2\pi} \ \theta_n \, Q}\exp\left[-\frac{(\ln Q-m_n)^2}{2\theta^2_n}\right],
\end{equation} 
where 
\begin{eqnarray}
m_n &=& m+\frac{\theta^2}{2}+\ln\left(\frac{n}{\sqrt{1+\frac{\exp(\theta^2)-1}{n}}}\right)\label{mn},\\
\theta_n &=& \sqrt{\ln\left(1+\frac{\exp(\theta^2)-1}{n}\right)}. \label{thetan}
\end{eqnarray}
Therefore, the distribution function of $Q$ for a given value of the average number of muons $\mu$ is given by
\begin{equation}
\label{convdist}
f_{\textrm{{\scriptsize ADC}}}(Q;\mu) \cong \left[ \delta(Q) + \sum_{n=1}^{\infty}%
\frac{\mu^n}{n!} g_{\textrm{{\scriptsize ADC}}}(Q;n) \right] \exp{(-\mu)}. 
\end{equation}
Here $\delta(Q)$ is the Dirac Delta function that accounts for the assumption of a constant, or noise-free, baseline of the ADC. 

Since the measured values of $k$ and $Q$ depend on the number of muons that hit a given muon detector, the joint distribution 
function of these two random variables can be written as
\begin{eqnarray}
f(k,Q;\mu)\! \! \! \! &=& \! \! \! \! \left[ \delta(Q) + %
\sum_{n=1}^{\infty} g_{\textrm{{\scriptsize ADC}}}(Q;n) \, P(k;n) \, \frac{\mu^n}{n!} \right] \times \nonumber \\
&& \exp{(-\mu)},
\label{fkQ}
\end{eqnarray}
where $P(k;n)$ is the probability of $k$ given $n$ which is given by \cite{Supanitsky:20}
\begin{equation}
P(k;n) = \binom{n_s}{k} \, S(n,k) \, \frac{k!}{n_s^n}.
\end{equation}
Here $S(n,k)$ is the Stirling number of second kind which is given by
\begin{equation}
S(n,k)=\frac{1}{k!} \, \sum_{j=0}^{k} \binom{k}{j} \, (-1)^j \, (k-j)^n. 
\end{equation}
For practical purposes, an approximate expression for $S(n, k)$ is used when $n \geq 70$. This approximation, 
along with an analysis of the associated error for $n \geq 70$, is provided in \ref{appS2}.

It is worth mentioning that to obtain Eq.~(\ref{fkQ}) it is assumed that $k$ and $Q$ are independent random 
variables, which is a reasonable supposition given that the deposited charge does not depend on the distribution 
of the muons among the detector segments.

Finally, the combined likelihood employed for parameter estimation is expressed as
\begin{equation}
\label{LikNew}
\mathcal{L}(\mu;k,Q)=f(k,Q;\mu).
\end{equation}

\section{Estimation of the mean number of muons}
\label{sec:EstMu}

Given a measurement of $k$ and $Q$, the parameter $\mu$  can be estimated by maximizing the likelihood function 
associated with the binary mode, the ADC mode, or the combination of both as given in Eq.~(\ref{LikNew}). The 
likelihood functions for the binary and ADC modes used in the following analyses are, respectively, given by
\begin{eqnarray}
\label{LikB}
&&\mathcal{L}_{\textrm{{\scriptsize B}}}(\mu;k)=P_{\textrm{{\scriptsize B}}}(k;\mu), \\[0.15cm] 
\label{LikADC}
&&\mathcal{L}_{\textrm{{\scriptsize ADC}}}(\mu;Q)=f_{\textrm{{\scriptsize ADC}}}(Q;\mu). 
\end{eqnarray}

The maximization of the considered likelihood functions must be performed numerically, except in the case of the 
binary likelihood. In this case, the maximum likelihood estimator of $\mu$ is given by \cite{Ravignani:15}
\begin{equation}
\label{muhbin}
\hat{\mu} = -n_{\textrm{s}} \ln\left[ 1-\frac{k}{n_{\textrm{s}}} \right].
\end{equation}

\begin{figure}[!ht]
\centering
\includegraphics[width=7.8cm]{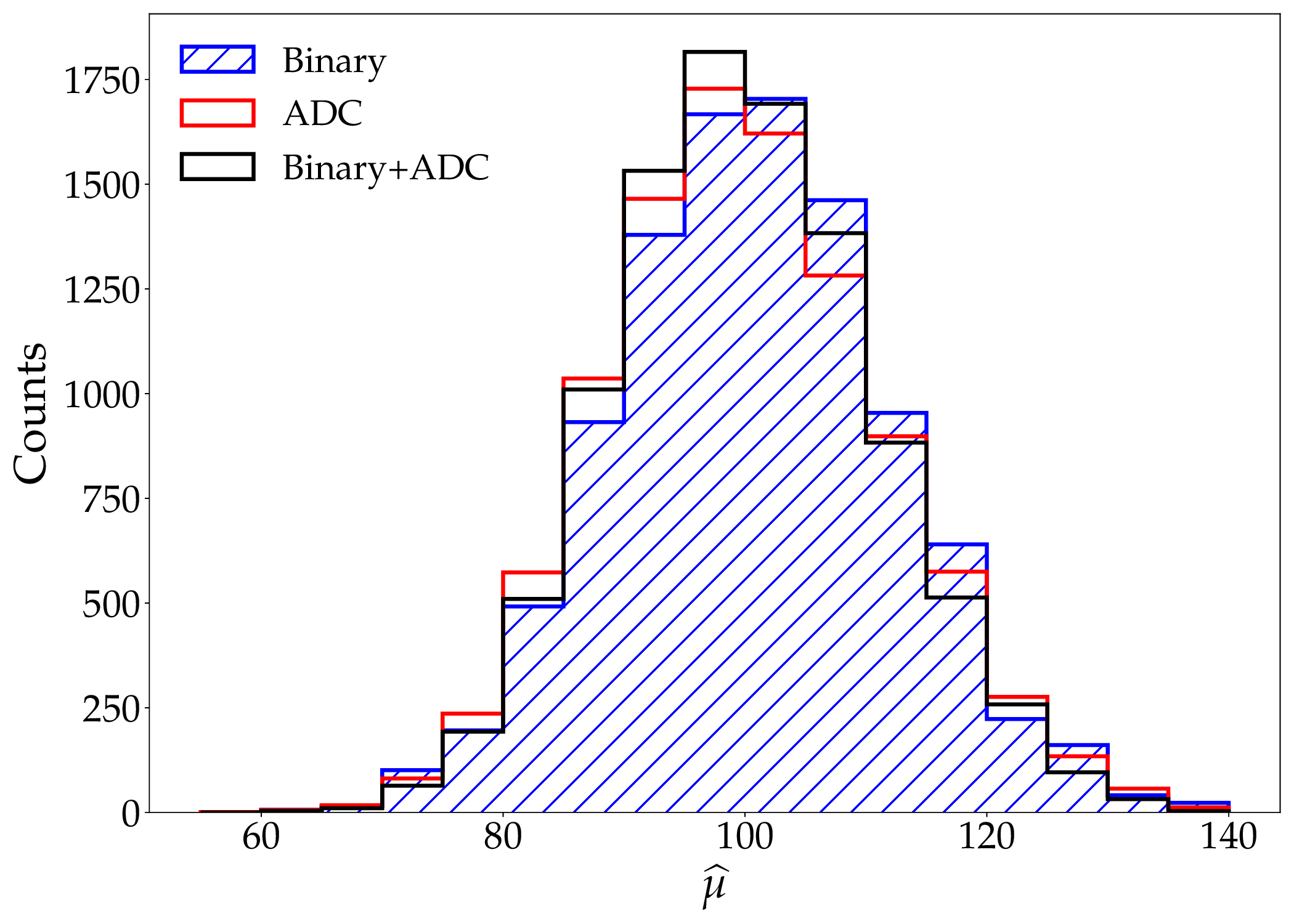}
\caption{Distributions of $\hat{\mu}$ obtained by maximizing the likelihood functions of the binary, the ADC and 
the combined one. The parameters used are: $\mu=100$ and $n_\textrm{s}=192$. \label{MuDist}}
\label{Muhat100}
\end{figure}

The performance of the different methods is assessed using a simplified detector simulation, described in 
Ref.~\cite{Varada:23}. This simulation incorporates the detector response for both binary and ADC acquisition 
modes, including the pile-up effect relevant to the binary mode and the saturation of both acquisition modes. 
For each value of $\mu$, the number of muons impacting a detector is determined by Poisson sampling. Correspondingly, 
for each incident muon count, a sample of $k$ and $Q$ values is generated. In total, $10^4$ samples of $k$ and $Q$ 
pairs are obtained per $\mu$ value, which are then used to estimate $\mu$ by maximizing the three likelihood 
functions considered, which is done by using the MINUIT \cite{Minuit} routine implemented in the ROOT package
\cite{root}. Figure \ref{MuDist} presents the $\hat{\mu}$ distributions for $\mu=100$ and $n_\textrm{s}=192$
(the segmentation of the Auger underground muon detectors), derived from these three likelihoods. 

To compare the performance of the three $\mu$ estimation methods, three quantities are evaluated for each $\mu$ 
value and reconstruction method: the relative bias ($\langle \hat{\mu} \rangle/\mu-1$), the relative standard 
deviation ($\sigma[\hat{\mu}]/\mu$), and the coverage probability of the confidence interval of $\mu$. As 
previously mentioned, for the binary mode, the maximum likelihood estimator is given by Eq.~(\ref{muhbin}). 
Both $\sigma[\hat{\mu}]$ and the coverage probability can also be calculated analytically for the binary case. 
For the ADC and combined methods, $\langle \hat{\mu} \rangle$ and $\sigma[\hat{\mu}]$ are derived from the 
$\hat{\mu}$ distributions obtained through simulations (see Fig.~\ref{Muhat100}). The coverage probability 
for these methods is determined by considering the uncertainty of each $\hat{\mu}$ value, which is derived 
considering a second order approximation of $-\ln \mathcal{L}$ around the estimated value of $\mu$ \cite{Minuit}.

\begin{figure}[!ht]
\centering
\includegraphics[width=7.8cm]{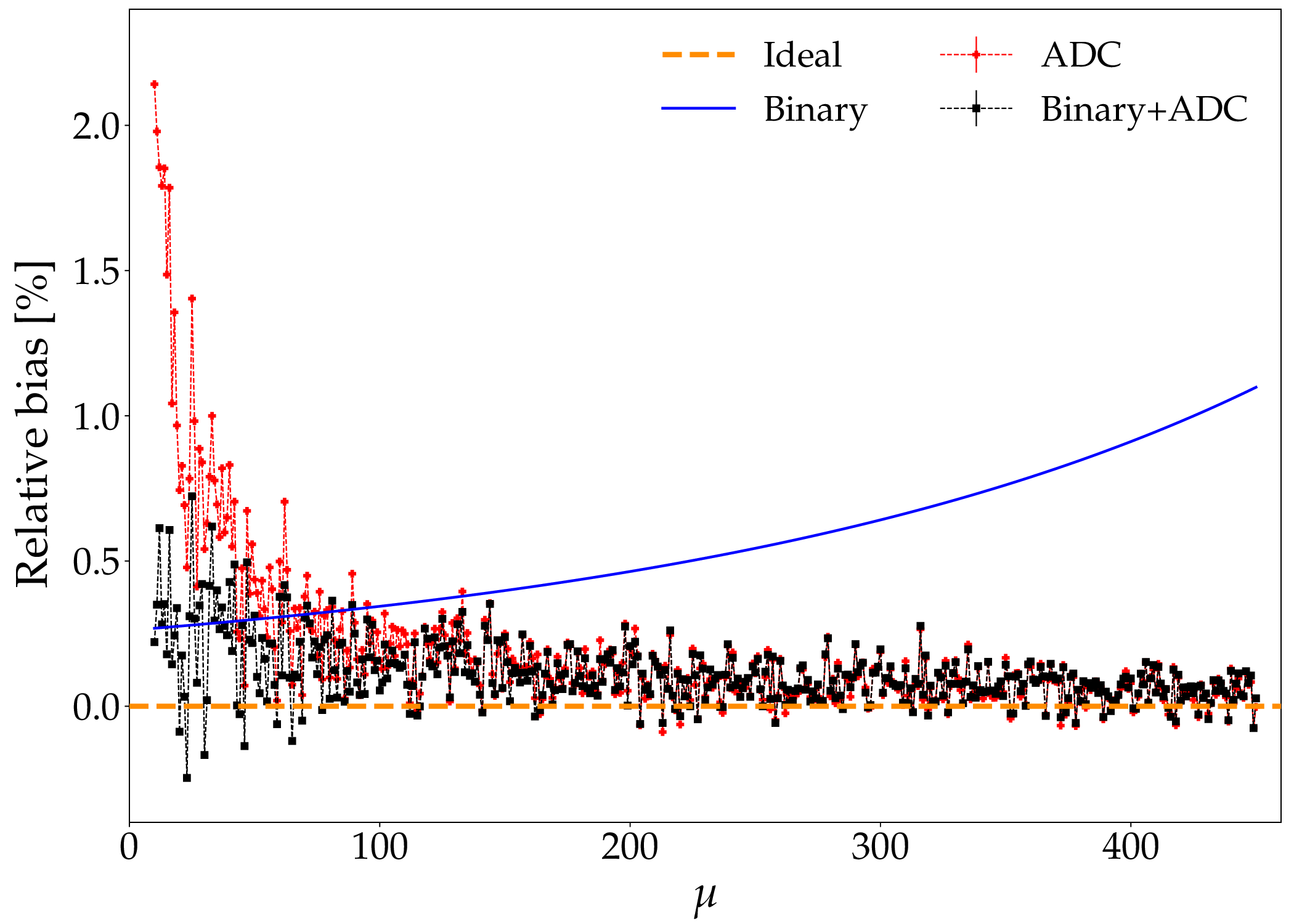}
\includegraphics[width=7.8cm]{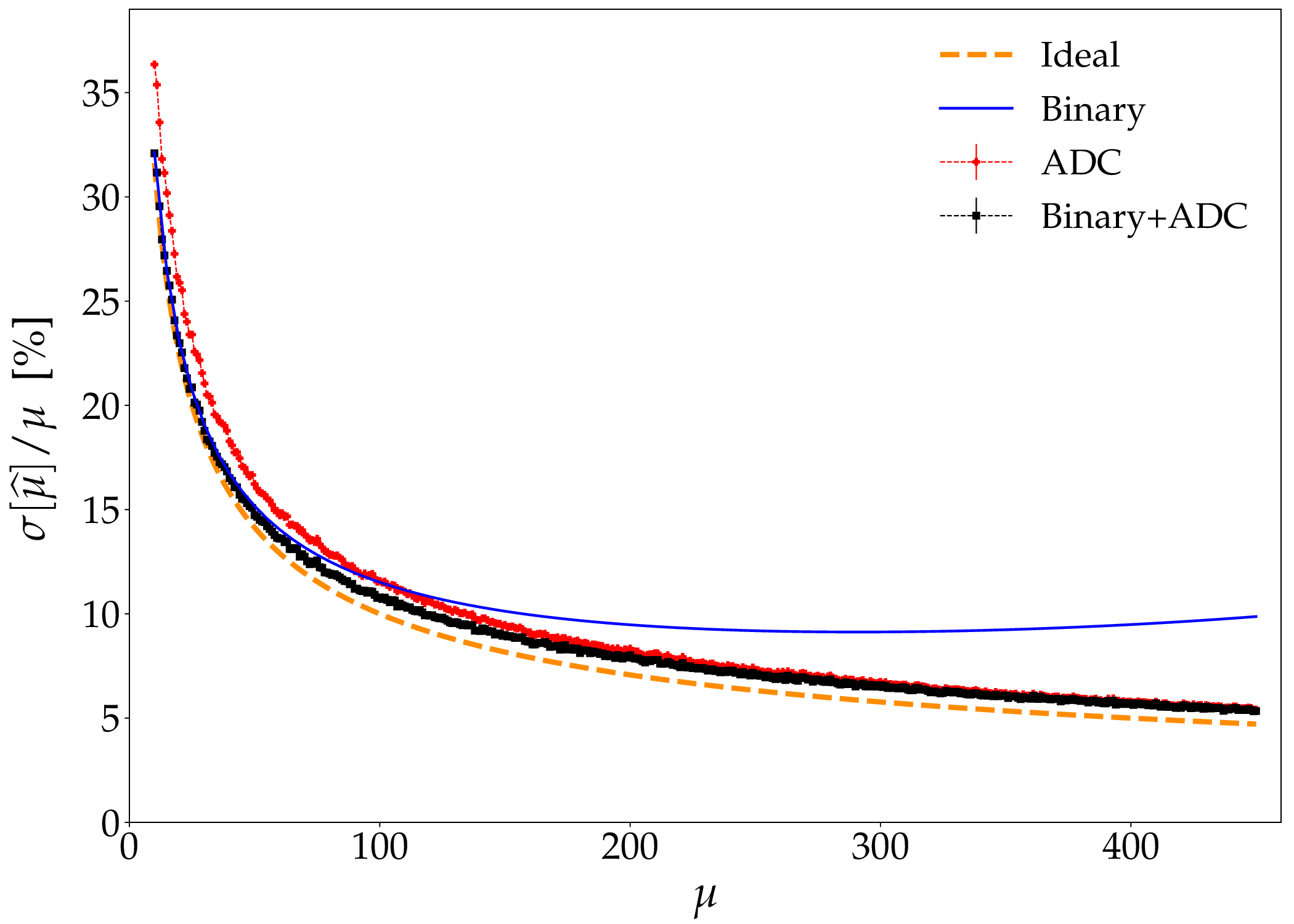}
\includegraphics[width=7.8cm]{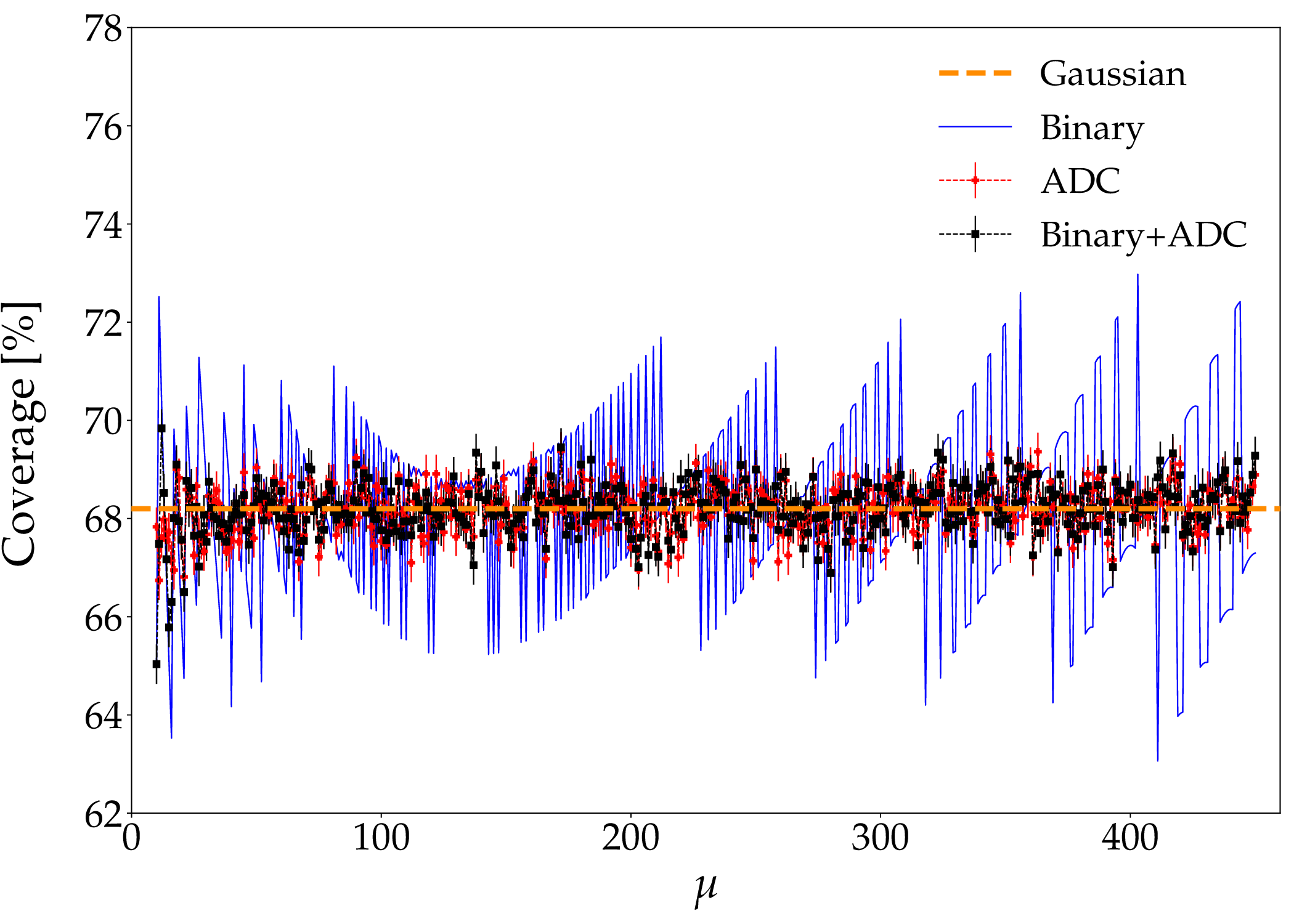}
\caption{Relative bias (top panel), relative standard deviation (middle panel) and coverage (bottom panel) of $\hat{\mu}$ 
as a function of $\mu$ obtained for the three methods considered and for $n_\textrm{s}=192$. For the relative bias and 
relative standard deviation the case of an ideal detector is also included. For the coverage the Gaussian case is also 
included.}
\label{BiasSigCov}
\end{figure}

The top panel of Fig.~\ref{BiasSigCov} shows the relative bias as a function of $\mu$ for $n_\textrm{s}=192$, 
obtained using the three methods considered and the result corresponding to an ideal detector for which
the likelihood function corresponds to the Poisson distribution \cite{Varada:23}. The relative bias of the 
binary mode is smaller than that of the ADC mode for small values of $\mu$. As $\mu$ increases, the relative 
bias associated with the ADC mode decreases, while that of the binary mode increases. Consequently, for large 
values of $\mu$, the relative bias of the ADC mode becomes smaller than that of the binary mode. Note that the 
increase of the bias of the binary mode arises from pile-up and saturation effects, with the latter becoming more 
significant at high incident muon counts. When using the binary mode alone, this bias can be corrected by applying 
an appropriate unbiased method. However, such a correction must be approximated, since the bias is not a linear 
function of $\mu$. The relative bias for the combined method remains smaller than, or at least comparable to, those 
of the binary and ADC methods across the entire range of $\mu$, being close to that of the binary mode at low $\mu$ 
and to that of the ADC mode at high $\mu$. The relative bias remains small for all three methods considered, with 
the bias of the combined method being less than $\sim 0.8\%$ for $n_\textrm{s}=192$ throughout the full range of 
$\mu$ values examined. It is worth mentioning that, for the ideal detector the maximum likelihood estimator is 
given by the number of impinging muons $n$. Since $\langle n \rangle = \mu$, the bias is zero.

\begin{figure}[!ht]
\centering
\includegraphics[width=7.8cm]{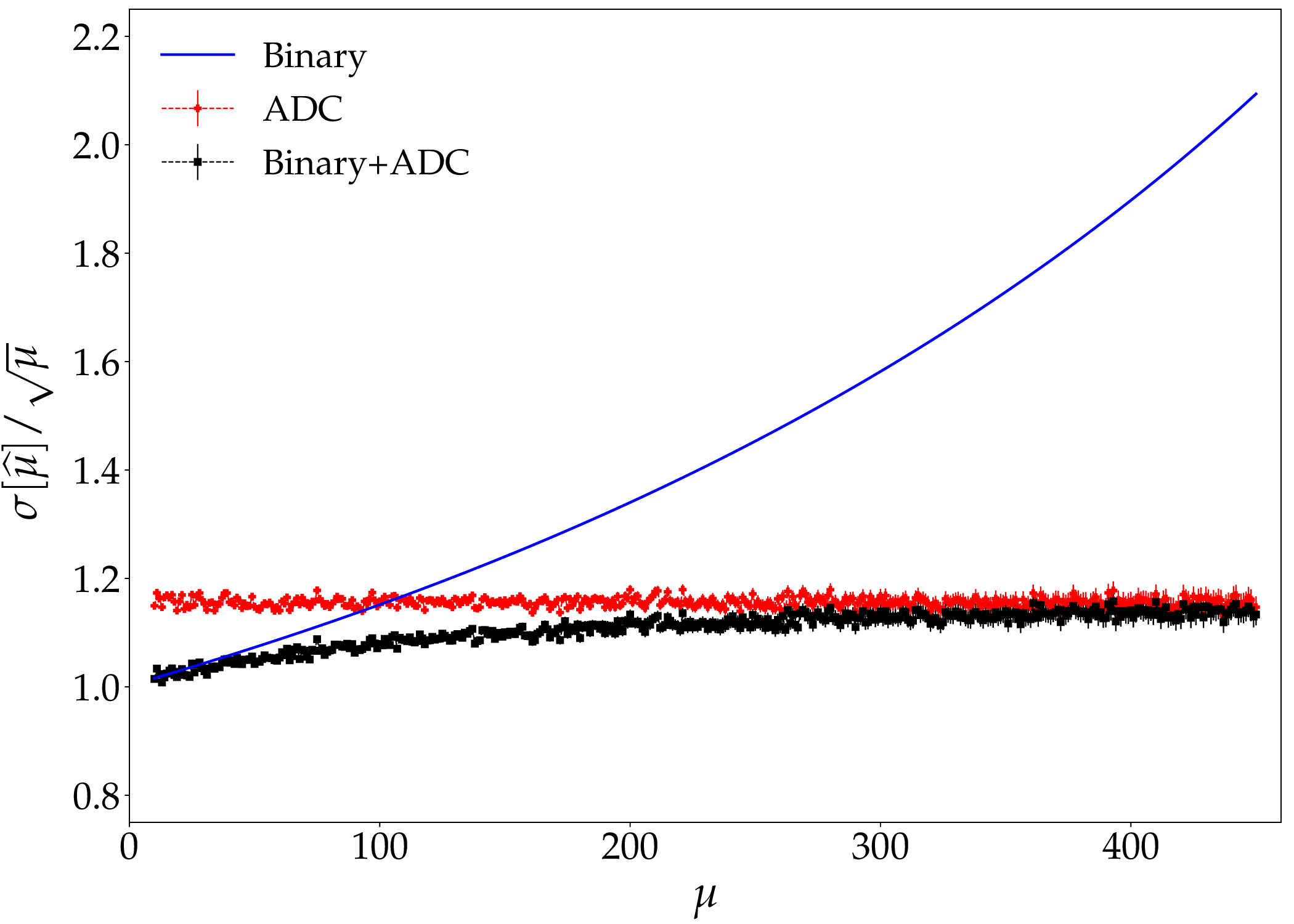}
\caption{Standard deviation of $\hat{\mu}$ relative to that of an ideal detector, $\sqrt{\mu}$, as a function 
of $\mu$ obtained for the three methods considered and for $n_\textrm{s}=192$.}
\label{SigPois}
\end{figure}

The middle panel of Fig.~\ref{BiasSigCov} shows the relative standard deviation as a function of $\mu$ for 
$n_\textrm{s}=192$, obtained using the three reconstruction methods, together with the result for an ideal detector. 
Similar to the behavior observed for the relative bias, the relative standard deviation of the binary mode is smaller 
than that of the ADC mode at low $\mu$, while at high $\mu$ the situation is reversed, with the ADC mode yielding 
smaller fluctuations. Across the full range of $\mu$, the combined likelihood consistently achieves a relative 
standard deviation that is lower than, or at least comparable to, those of the individual binary and ADC modes, 
approaching the binary mode at low $\mu$ and the ADC mode at high $\mu$. For an ideal detector, the expected relative 
standard deviation is $1/\! \sqrt{\mu}$. As shown in the figure, the performance of the combined method closely follows 
this ideal behavior. This is illustrated in more detail in Fig.~\ref{SigPois}, which shows the ratio 
$\sigma[\hat{\mu}]/\! \sqrt{\mu}$, where $\sqrt{\mu}$ is the standard deviation for an ideal detector. At small $\mu$, 
where the combined likelihood is dominated by the binary contribution, $\sigma[\hat{\mu}]$ is nearly indistinguishable 
from the ideal value. This occurs because, in this regime, the number of detector segments is much larger than the number 
of incident muons, and then the pile-up effect is negligible. As $\mu$ increases, the ratio $\sigma[\hat{\mu}]/\! \sqrt{\mu}$ 
for the combined method grows moderately, but the deviation from the ideal detector remains below $\sim 15 \%$ over the 
entire $\mu$ range considered. For the ADC mode, this ratio decreases slowly with increasing $\mu$, ranging from about 
$1.18$ to $1.15$. In contrast, the binary mode matches the ideal case at low $\mu$, as expected, but its ratio rises 
rapidly with increasing $\mu$, reaching values close to 2.1 at $\mu = 450$.

From the top and middle panels of Fig.~\ref{BiasSigCov}, it appears that the combined likelihood function is dominated 
by the binary mode at small values of $\mu$, exhibits a transition region at intermediate values of $\mu$, and becomes 
dominated by the ADC mode at large values of $\mu$. The transition region seems to be centered around $\mu \sim 100$. In 
this transition region, both the relative bias and the relative standard deviation for the combined likelihood are smaller 
than those obtained using the binary and ADC modes individually.

The bottom panel of Fig.~\ref{BiasSigCov} shows the coverage of $\hat{\mu}$ as a function of $\mu$ obtained for the three 
methods considered. As observed in the figure, the coverage for the ADC and combined methods is quite close to that expected 
for a Gaussian likelihood. In contrast, the coverage for the binary mode exhibits a different behavior, deviating significantly 
from the Gaussian expectation at specific values of $\mu$. This deviation arises from the discrete nature of $k$.

\begin{figure}[!ht]
\centering
\includegraphics[width=7.8cm]{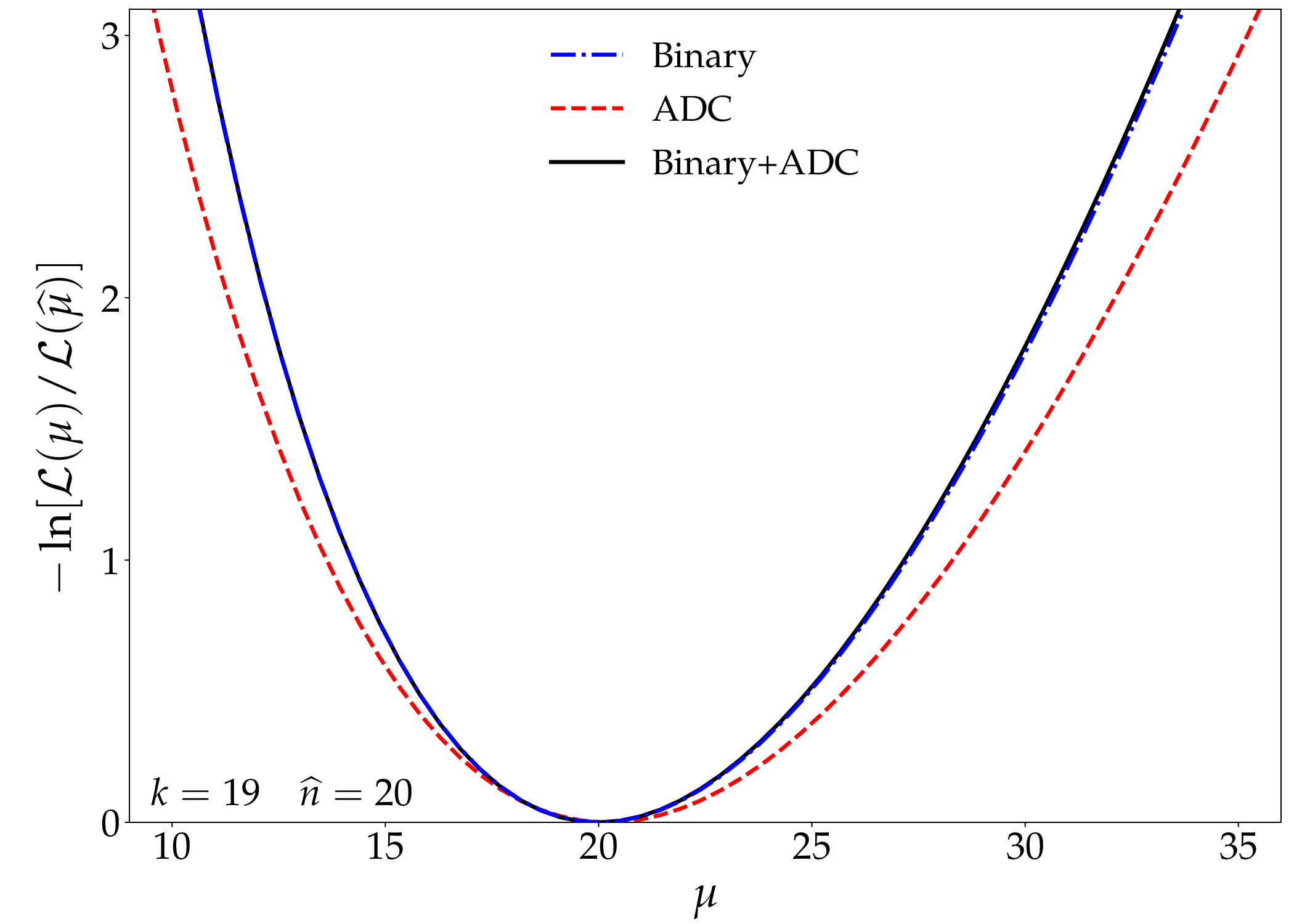}
\includegraphics[width=7.8cm]{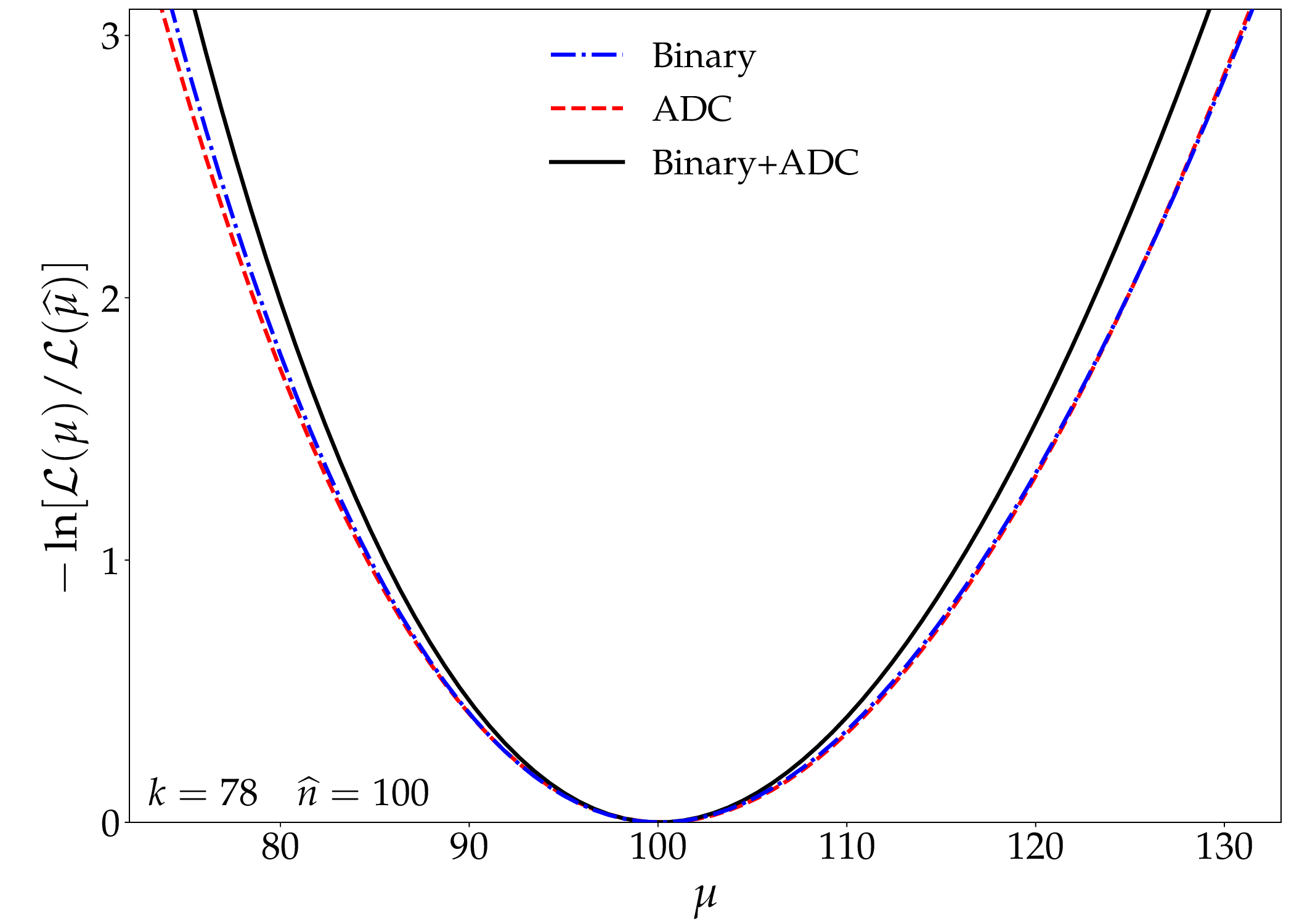}
\includegraphics[width=7.8cm]{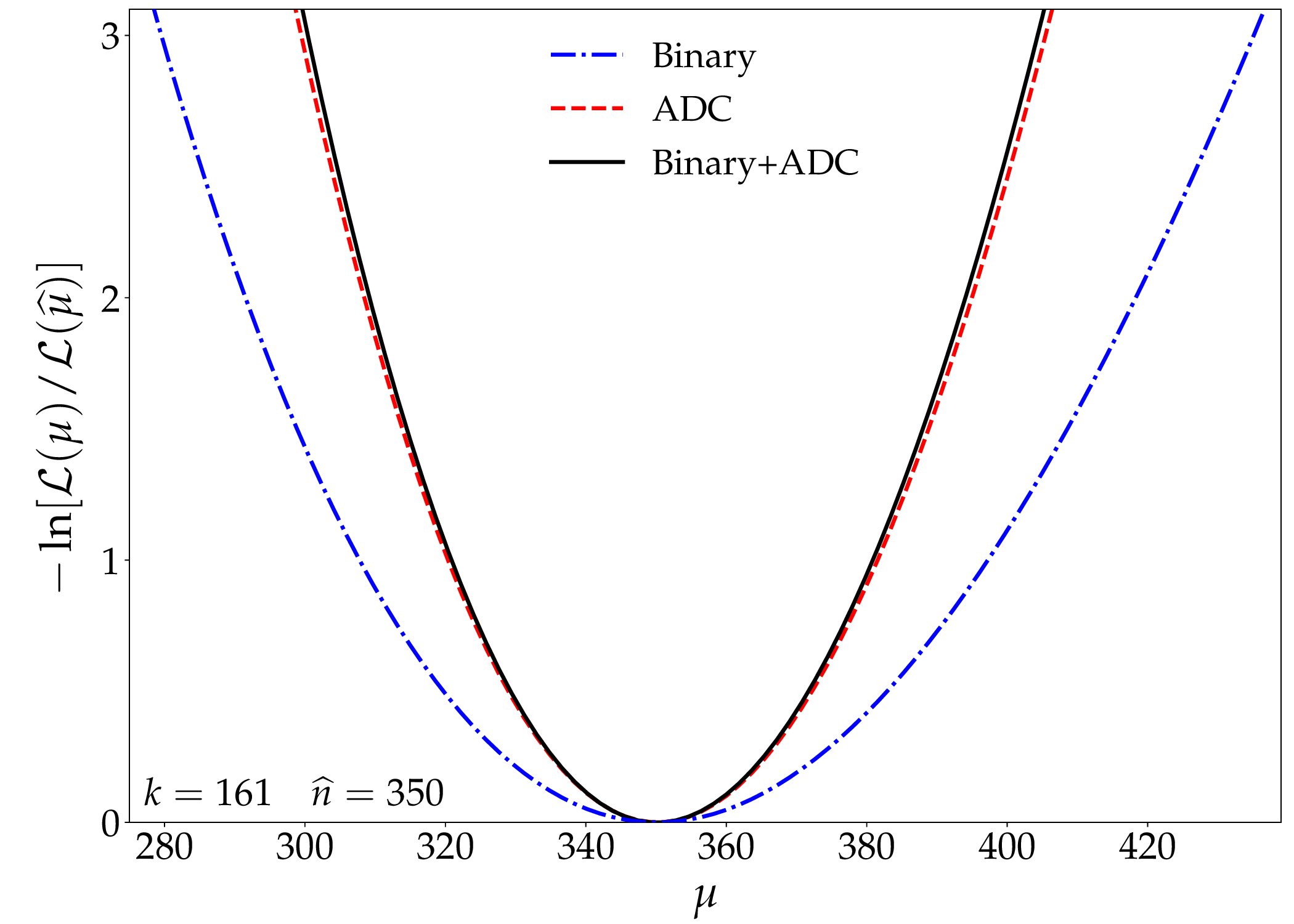}
\caption{The negative logarithm of the likelihood ratio as a function of $\mu$ obtained using the three methods considered. 
Each plot corresponds to a different region of the parameter $\mu$. The number of segments used is $n_\textrm{s}=192$.}
\label{Likelihood}
\end{figure}

The performance of the $\mu$ estimator for the different methods considered is closely related to the shape of the 
corresponding likelihood function. Figure \ref{Likelihood} shows the negative logarithm of the likelihood ratio as a function 
of $\mu$ for the three methods considered, and for different values of $k$ and $\hat{n} = Q / \langle q \rangle$, where 
$\langle q \rangle$ is the mean charge deposited by a single muon. From the top panel of the figure, which corresponds to 
small values of $\mu$, it can be seen that the likelihood functions of the binary and combined methods are very similar, 
while the likelihood corresponding to the ADC method is broader than the other two. In the middle panel, corresponding to 
intermediate values of $\mu$, the likelihood functions of the binary and ADC methods are similar, and the one associated 
with the combined method is narrower. From the bottom panel, which corresponds to large values of $\mu$ (i.e., on the order 
of 350), it is evident that the likelihood function of the binary method is significantly broader than those of the other 
two. In this case, the combined likelihood remains slightly narrower than that of the ADC method.

\section{Reconstruction of the muon lateral distribution function}
\label{sec:Rec}

The function used to fit the simulated data is given by a KASCADE-Grande-like MLDF \cite{KG:10}, which can be written as follows    
\begin{equation}
\mu(r;\boldsymbol{p})  = \mu_0 \frac{h(r;\beta)}{h(r_0;\beta)}.
\label{KGLMLDF}
\end{equation}
Here $\boldsymbol{p}=(\mu_0, \beta)$ where $\mu_0$ is a normalization factor and $\beta$ an exponent of the shape 
function  
\begin{equation}
h(r;\beta) = \left(\frac{r}{r_1} \right)^{-\alpha}\left(1+\frac{r}{r_1} \right)^{-\beta}\left( 1+\left( \frac{r}{10 \, r_1}\right)^2 \right)^{-\gamma}.
\end{equation}
The assumed values of the fixed parameters are: $\alpha=0.75$, $r_1=320$ m, $\gamma=4.18$ (see Ref.~\cite{Varada:23}), 
and $r_0 = 450$ m, which is a reference distance for the 750 m array of Auger \cite{Varada:23}. The fitted MLDF evaluated 
at the distance $r_0$, i.e.~$\hat{\mu}(450)\equiv \mu(r_0=450\, \textrm{m};\hat{\boldsymbol{p}})$, is used to measure the 
muon size of the shower.

The stations that participate in a given event are classified as triggered (T), non-triggered (NT), and 
saturated (S) and different expressions of the likelihood function are used in each case. The free-fit 
parameters in $\boldsymbol{p}$ are obtained by minimizing the function
\begin{equation}
-2 \ln \mathcal{L} = -2 \sum_{i=1}^{N} \ln \left[\mathcal{L}_i(\mu(r_i,\boldsymbol{p})) \right]. 
\label{CI}
\end{equation}
where the sum runs over the $N$ stations participating in the event reconstruction and the likelihood $\mathcal{L}_i$ varies 
according to the station classification described above. 

Non-triggered stations are defined as those for which $k \leq 2$ \cite{Ravignani:15}. In this case, the likelihood function 
used corresponds to the binary mode \cite{Ravignani:15}, as it performs better than that of the ADC mode for small values 
of impinging muons. Its expression is 
\begin{eqnarray}
\mathcal{L}_\textrm{NT}(\mu) &=&  \exp(-\mu) \left[ 1+ n_{\textrm{s}} \, (\exp(\mu/n_{\textrm{s}})-1) + \right. \nonumber \\[0.4cm]
&& \left. \frac{n_{\textrm{s}}(n_{\textrm{s}}-1)}{2} \, (\exp(\mu/n_{\textrm{s}})-1)^2 \right].
\end{eqnarray}

The likelihood function for triggered stations in which none of the modes are saturated is given by Eq.~(\ref{LikNew}). 
However, in Sec.~\ref{sec:EstMu} it was shown that, for larger values of $Q$, the combined likelihood resembles the 
likelihood of the ADC mode, typically for $Q \gtrsim 350 \, \langle q \rangle$ (see lower panel of Fig.~\ref{Likelihood}). 
Moreover, Ref.~\cite{Varada:23} demonstrates that the ADC likelihood can be safely approximated by a Gaussian likelihood 
when $Q\geq 200 \, \langle q \rangle$. Therefore, for large values of $Q$ the likelihood is approximated by a Gaussian to 
reduce the computation time during the minimization procedure. The likelihood corresponding to the triggered stations for 
the case in which none of the modes is saturated is given by
%
%
%
%
%
%
%
%
%
%
%
%
%
%

\begin{equation}
\mathcal{L}_\textrm{T}(\mu;k,Q) =  
\left\{ 
\begin{array}{ll}
\! \!  \! f(k,Q; \mu) & \! \! k>2 \, \& \, Q \in (0,350 \, \langle q \rangle) \\[0.2cm]
%
\! \! \! \mathcal{N}(Q;\mu_Q,\sigma^2_Q) & \! \! Q \geq 350 \, \langle q \rangle
\end{array}
\right. 
\label{LNT}
\end{equation}      
where $\mathcal{N}(Q;\mu_Q,\sigma^2_Q)$ is the normal distribution with the mean value of $Q$ and its standard 
deviation given by \cite{Varada:23}
\begin{eqnarray}
\mu_Q &=& \mu \, \langle q \rangle \\[0.2cm]
\sigma^2_Q &=& \mu \, \left (\varepsilon^2[q]+1 \right ) \langle q \rangle^2.
\end{eqnarray} 
Here $\varepsilon[q]=\sigma[q]/\langle q \rangle$ which for the Auger underground muon detectors takes the value 
$\varepsilon[q]\cong 0.6$ \cite{Botti:19}.

Since the two acquisition modes operate independently, it is possible for a given station to experience saturation 
in one mode but not in the other. In most cases, saturation in the ADC mode coincides with saturation in the binary 
mode. However, in approximately $1\%$ of cases, the ADC mode is saturated while the binary mode remains unsaturated. 
Therefore, in cases where the ADC mode is not saturated but the binary mode is, the Gaussian likelihood of Eq.~(\ref{LNT}) 
is used, since the condition $Q \geq 350\, \langle q \rangle$ is always satisfied. When the ADC mode is saturated and 
the binary mode is not, the likelihood function corresponding to the binary mode (see Eq.~(\ref{LikB})) is used.

A station is flagged as saturated when both acquisition modes are saturated. In such cases, since the likelihood 
is dominated by the ADC at high values of $Q$, the expression corresponding to saturated stations from the ADC-based 
reconstruction method is applied \cite{Varada:23}
\begin{equation}
\mathcal{L}_{\textrm{S}}(\mu;Q) = \frac{1}{2} \left( 1-\textrm{erf} \left[ \frac{Q-\langle Q \rangle}{\sqrt{2}\, \sigma[Q]} \right] \right),
\end{equation}
where $Q$ is the value of the charge measured by the ADC obtained integrating the saturated pulse and $\textrm{erf}(x)$ 
is the error function.  

The performance of the new method to reconstruct the MLDF is studied from simulations. As mentioned before, the simulation 
program updated with the simplified simulation of the ADC described in Ref.~\cite{Varada:23} is used. The new reconstruction 
method is also implemented in the program. As described in Ref.~\cite{Varada:23}, for a given primary type, energy, and zenith 
angle, the program takes as input a MLDF obtained by fitting the average MLDF from simulated air showers with a 
KASCADE-Grande-like function (see Eq.~(\ref{KGLMLDF})), together with the average muon time distribution as a function of 
the distance to the shower axis. Both inputs are derived from the shower library of Ref.~\cite{Varada:23}, which was generated 
using CORSIKA v7.7100 \cite{Heck:97} with EPOS-LHC \cite{Tanguy:1} as the high-energy hadronic interaction model and FLUKA 
\cite{Fluka:1, Fluka:2} as the low-energy interaction model. The library includes proton and iron showers with energies in 
the range $\log_{10}(E/\textrm{eV}) \in [17.5, 19]$, in steps of $\Delta \log_{10}(E/\textrm{eV}) = 0.25$, for zenith angles 
of $30^\circ$ and $45^\circ$, and with uniformly distributed azimuth angles between $-180^\circ$ and $180^\circ$.

For each combination of primary type, energy, zenith angle, and reconstruction method, $10^4$ events are simulated and 
reconstructed. The shower direction and core position are assumed to be determined by an independent detector system, as 
in Auger, where both observables are reconstructed using data from the water Cherenkov detectors. In the present analysis, 
uncertainties in the determination of the shower axis and core position are not taken into account. The reconstruction 
methods evaluated are: $i)$ the combined method described above (Binary+ADC); $ii)$ the binary-mode method with an infinite 
time window (Binary), as presented in Ref.~\cite{Ravignani:15}, which employs the likelihood function given by Eq.~(\ref{LikB}); 
$iii)$ the ADC-mode method (ADC) described in Ref.~\cite{Varada:23}, based on the likelihood of Eq.~(\ref{LikADC}); and $iv)$ 
the ideal-detector method (Ideal), which assumes a Poisson likelihood to reconstruct events simulated with Poissonian muon 
detectors, i.e., detectors with infinite segmentation (see Ref.~\cite{Varada:23} for further details).

\subsection{Non-saturated events}

\begin{figure}[!ht]
\centering
\includegraphics[width=7.8cm]{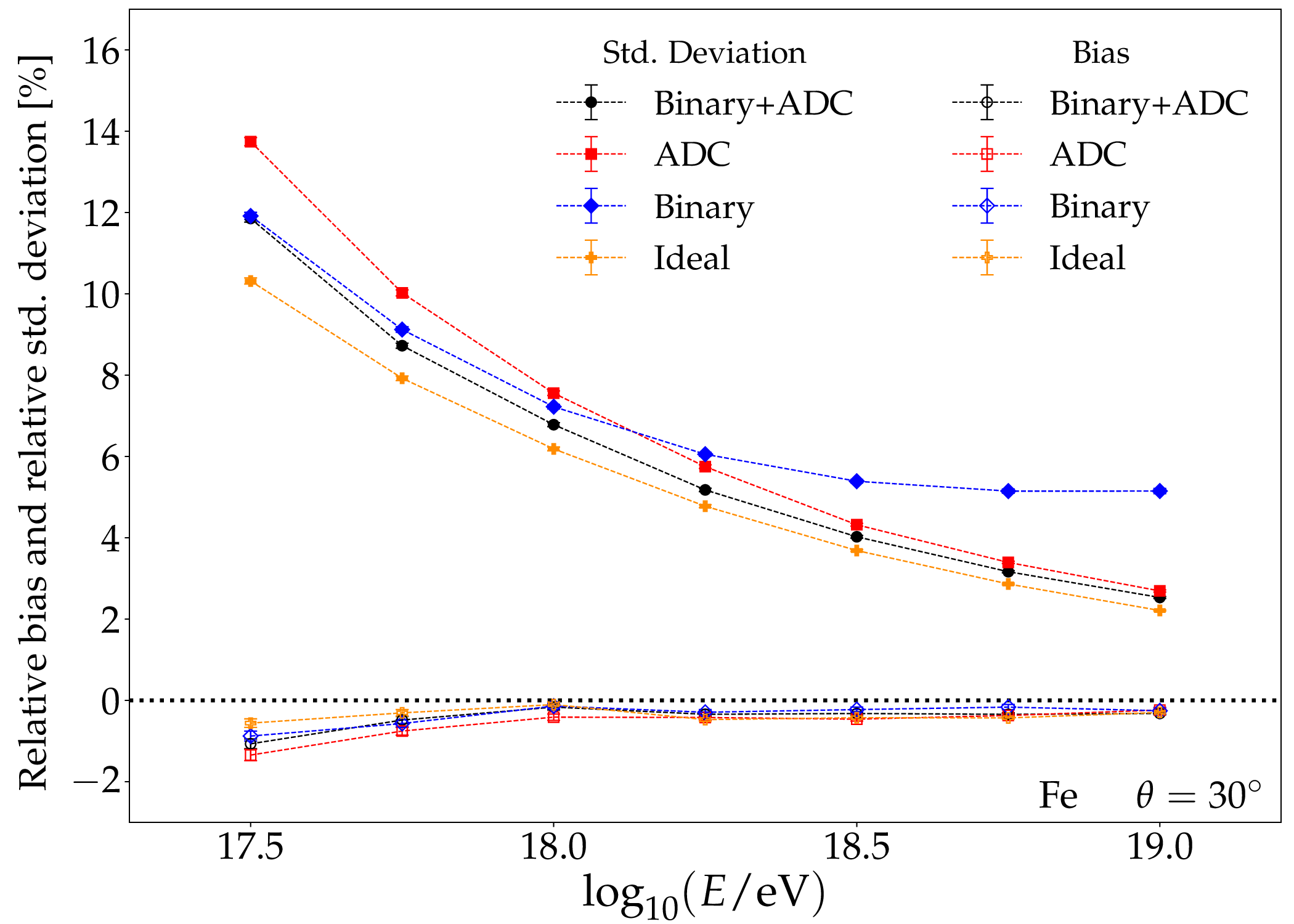}
\includegraphics[width=7.8cm]{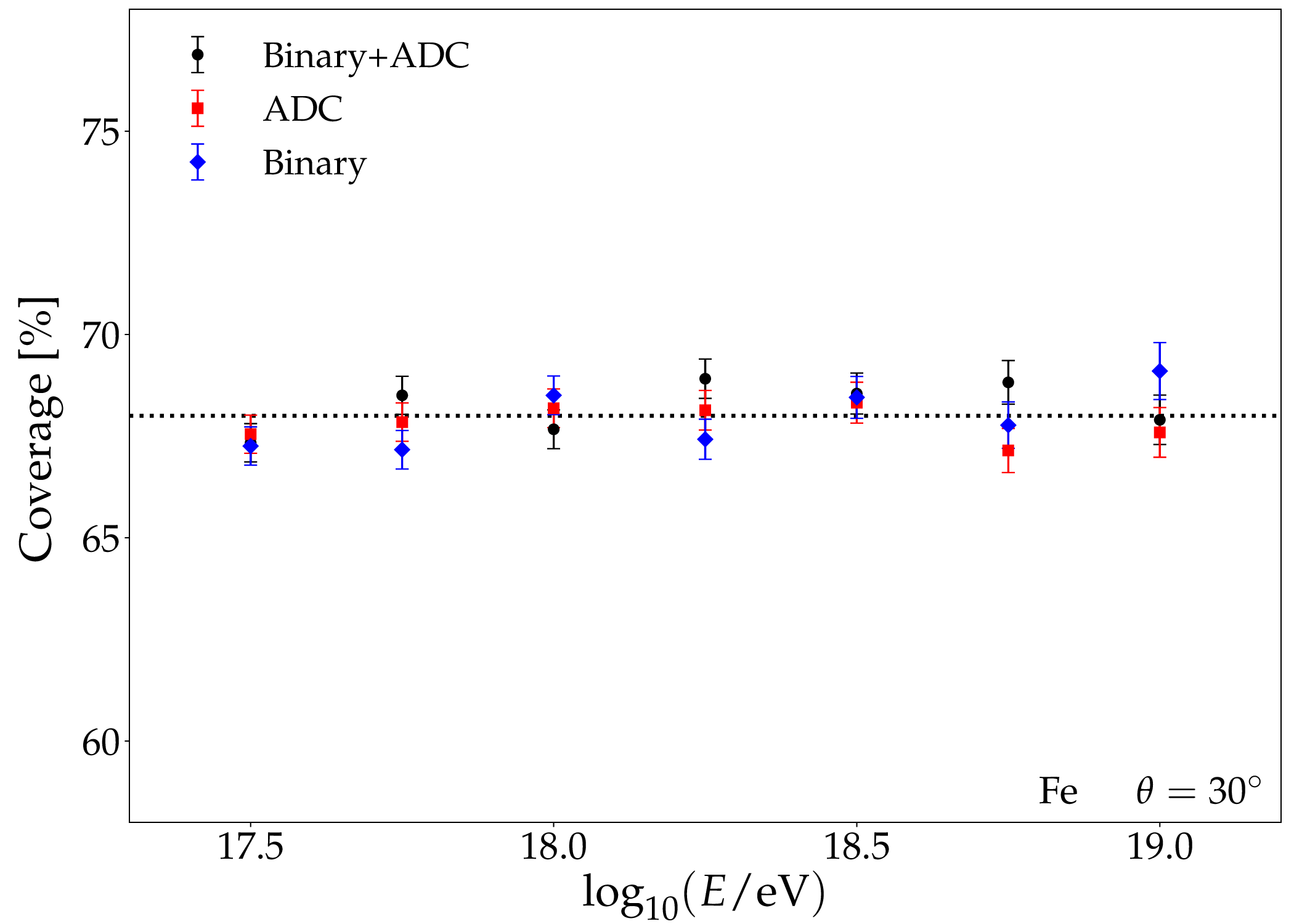}
\caption{Top panel: relative bias and relative standard deviation of $\hat{\mu}(450)$ as a function of the logarithm 
of the primary energy for the four reconstruction methods considered. Bottom panel: coverage of $\hat{\mu}(450)$ 
corresponding to the binary, ADC, and the combined method to reconstruct the MLDF. The dotted line corresponds to the 
coverage of a Gaussian likelihood. The reconstructed showers are generated by iron primaries of 30$^\circ$ zenith 
angle.}
\label{NonSatFe}
\end{figure}

\begin{figure}[!ht]
\centering
\includegraphics[width=7.8cm]{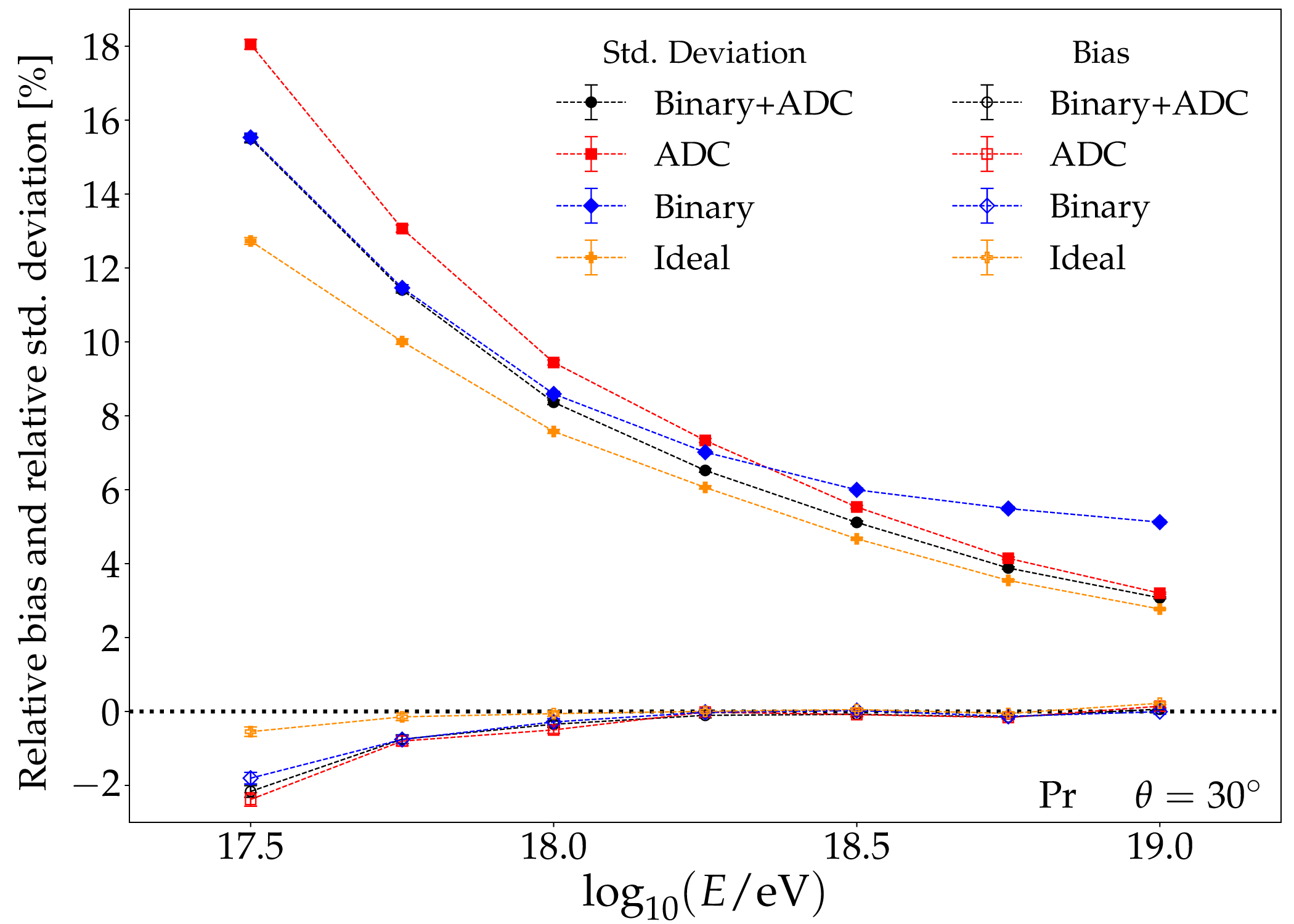}
\includegraphics[width=7.8cm]{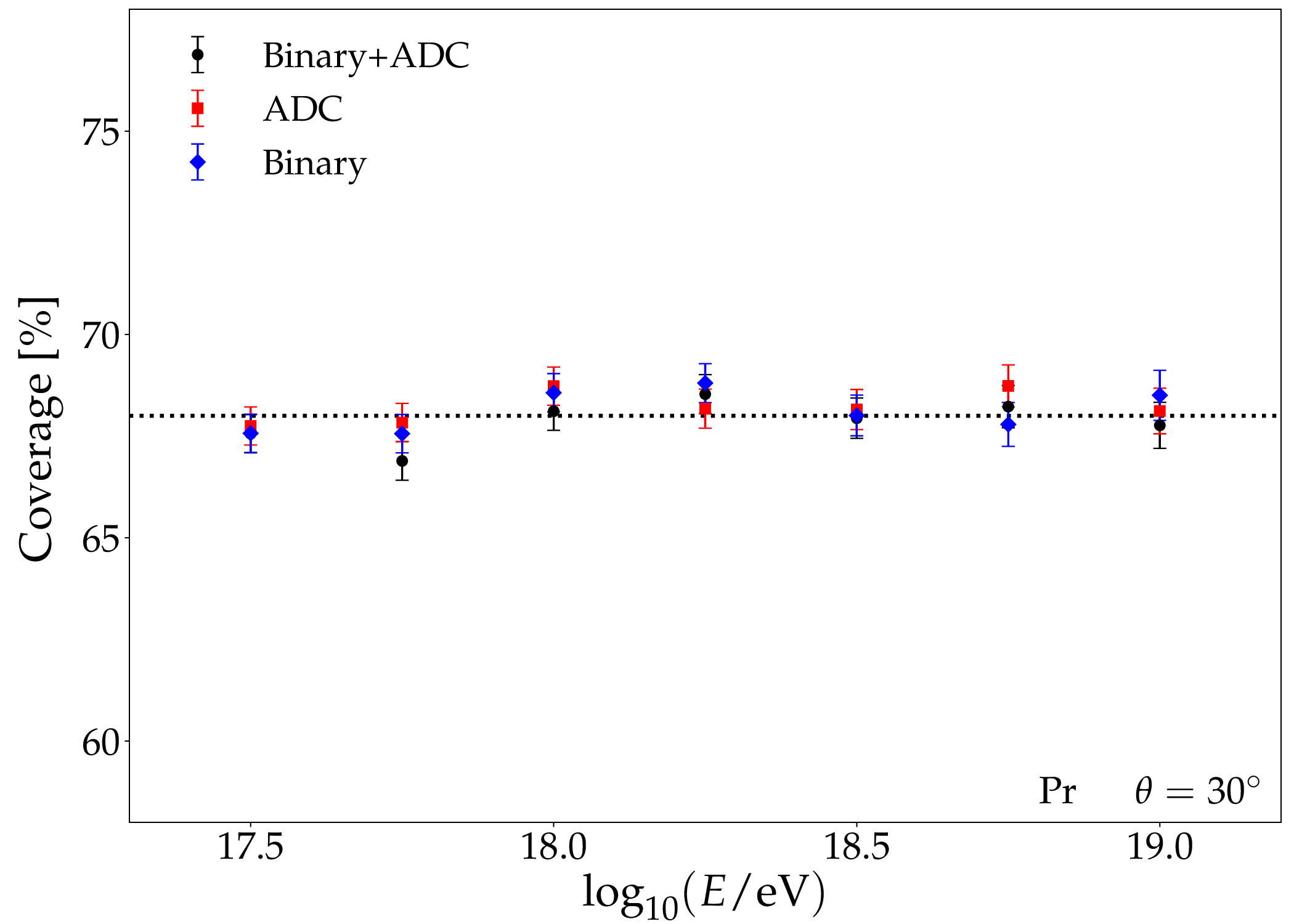}
\caption{Top panel: relative bias and relative standard deviation of $\hat{\mu}(450)$ as a function of the logarithm 
of the primary energy for the four reconstruction methods considered. Bottom panel: coverage of $\hat{\mu}(450)$ 
corresponding to the binary, ADC, and the combined method to reconstruct the MLDF. The dotted line corresponds to the 
coverage of a Gaussian likelihood. The reconstructed showers are generated by proton primaries of 30$^\circ$ zenith 
angle.}
\label{NonSatPr}
\end{figure}

The performance of the different reconstruction methods for the MLDF is evaluated using non-saturated events. Figures 
\ref{NonSatFe} and \ref{NonSatPr} show the relative bias and standard deviation of $\hat{\mu}(450)$ and the coverage 
of the confidence interval of $\mu(450)$ calculated from the likelihood of Eq.~(\ref{CI}). These quantities 
are shown as a function of the logarithm of the primary energy in Figs.~\ref{NonSatFe} and \ref{NonSatPr} for iron 
and proton showers, respectively. For each combination of primary particle type, energy, zenith angle, and reconstruction 
method, both the relative bias and the relative standard deviation of $\hat{\mu}(450)$ are determined by fitting the 
corresponding $\hat{\mu}(450)$ distributions with a Gaussian function. Note that the true value of $\mu(450)$ used in these
calculations is obtained by evaluating the input MLDF function at $r_0=450$ m.

As shown in the top panel of Fig.~\ref{NonSatFe}, the bias in $\hat{\mu}(450)$ remains small across the entire energy 
range considered. Additionally, the combined method yields a smaller relative standard deviation compared to those obtained 
using only binary- or ADC-based methods. Notably, at low energies, the relative standard deviation of the binary mode 
approaches that of the combined method; however, it increases significantly at higher energies. In contrast, the ADC-based 
method exhibits the opposite behavior; it performs worse at low energies but improves with increasing energy, eventually 
approaching the performance of the combined method. This confirms the expected behavior, namely, that the combined likelihood 
is dominated by the binary mode at low energies and by the ADC mode at high energies.

For iron primaries, the relative standard deviation of $\hat{\mu}(450)$ obtained using the combined method differs by less 
than $2\%$ from that of the ideal detector. The bottom panel of Fig.~\ref{NonSatFe} shows the coverage of $\hat{\mu}(450)$ 
for the different reconstruction methods. The results indicate that the coverage values obtained using the binary, ADC, and 
combined acquisition modes are all consistent with the expected value for a Gaussian likelihood.

Figure \ref{NonSatPr} presents the same analysis as Fig.~\ref{NonSatFe}, but for proton primaries. The overall behavior 
of the evaluated quantities is similar to that observed for iron showers. However, the relative standard deviation and the 
absolute value of the bias of $\hat{\mu}(450)$ are notably larger at low energies compared to the iron case. This difference 
arises because iron-induced showers produce more muons than proton-induced ones, thereby reducing the statistical uncertainties 
in the reconstruction of the MLDF. Despite this, the relative standard deviation of $\hat{\mu}(450)$ for proton primaries still 
reflects the same trend observed with iron; the combined likelihood is dominated by the binary mode at low energies and the 
ADC mode at high energies. Furthermore, for proton showers, the combined likelihood method continues to outperform both the 
binary-only and ADC-only methods, yielding a relative standard deviation of $\hat{\mu}(450)$ that is very close to that achieved 
by the ideal detector.

Consistent results are obtained for both proton and iron primaries at a zenith angle of 45$^\circ$. These findings 
confirm that the new combined reconstruction method outperforms those based solely on either the binary or ADC acquisition 
modes. Moreover, it demonstrates superior performance, particularly at low energies, compared to the combined method 
proposed in Ref.~\cite{Varada:23}. A key difference lies in the implementation of the likelihood function. In the present 
work, both acquisition modes are utilized simultaneously for nearly all detector stations. In contrast, the method described 
in Ref.~\cite{Varada:23} applies the binary likelihood only to stations registering a small number of muons, and the ADC 
likelihood to those with larger signals. In that approach, the transition between likelihood types is manually optimized 
based on performance metrics. The new method eliminates the need for such a transition criterion, offering a more seamless 
and robust treatment of the data.

\subsection{Saturated and non-saturated events}

\begin{figure}[!ht]
\centering
\includegraphics[width=7.8cm]{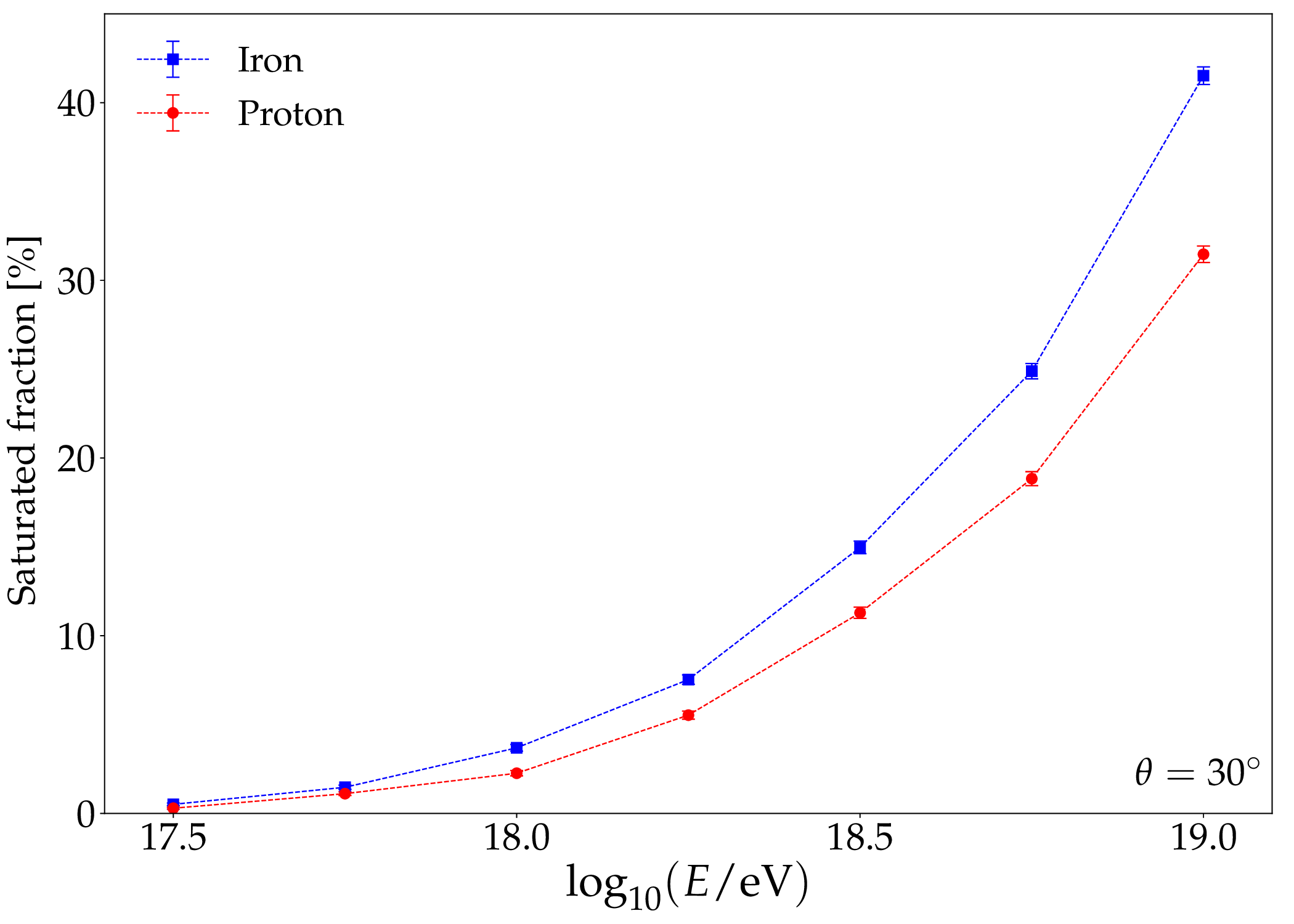}
\caption{Fraction of saturated events as a function of the logarithm of the primary energy for proton and iron
showers of 30$^\circ$ zenith angle.}
\label{SatFr}
\end{figure}

As previously mentioned, an event is considered saturated if it contains at least one saturated station. Figure \ref{SatFr} 
shows the fraction of saturated events as a function of primary energy for proton and iron primaries at a zenith angle of 
30$^\circ$. The figure indicates that the fraction of saturated events increases rapidly with energy, reaching approximately 
$30-40\%$ at $10^{19}$ eV. This fraction of saturated events is systematically higher for iron-induced showers due to their 
greater muon content compared to proton showers.

\begin{figure}[!ht]
\centering
\includegraphics[width=7.8cm]{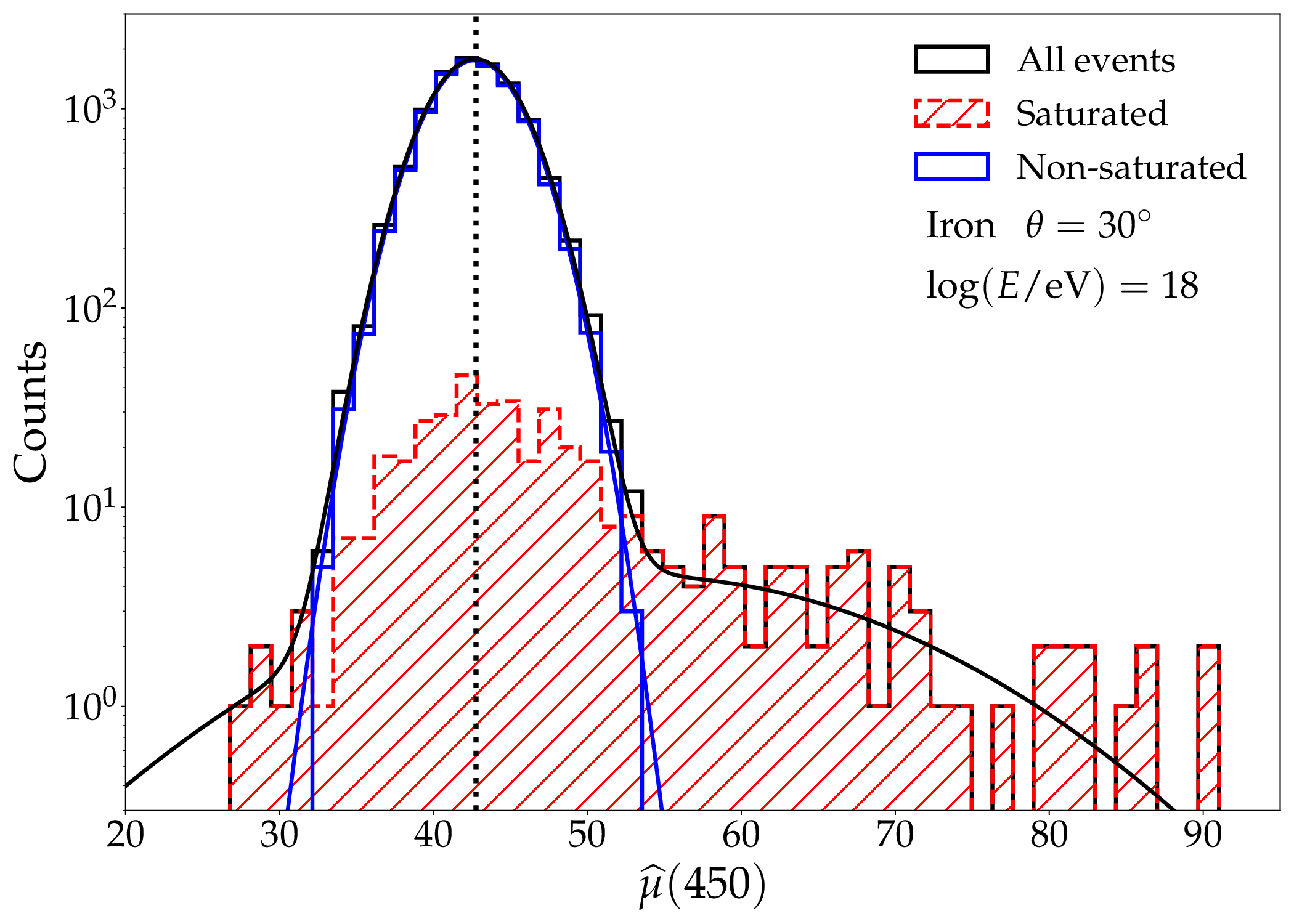}
\caption{Distributions of $\hat{\mu}(450)$ corresponding to non-saturated, saturated and all events for iron
showers of $10^{18}$ eV and 30$^\circ$ zenith angle. The blue solid line corresponds to the fit of the 
non-saturated histogram with a Gaussian function and the black solid line corresponds to the fit of the 
histogram of all events with double-Gaussian function. The dotted vertical line represent the true value
of $\mu(450)$.}
\label{SatDist}
\end{figure}

Saturated events are more challenging to reconstruct because saturated stations contribute to the likelihood function only 
as lower limits on the number of muons at their respective distances. Figure \ref{SatDist} presents the distributions of 
$\hat{\mu}(450)$, reconstructed using the combined method, for iron showers with a primary energy of $10^{18}$ eV and a 
zenith angle of $30^\circ$. The distributions are shown separately for non-saturated events, saturated events, and the 
full event set. As seen in the figure, the distribution corresponding to saturated events exhibits a pronounced tail 
extending toward larger values of $\hat{\mu}(450)$. Also shown are a Gaussian fit to the distribution of non-saturated 
events and a double-Gaussian fit applied to the distribution of all events, capturing the broader structure introduced 
by saturation effects.

As previously discussed, the muon content of air showers is a parameter highly sensitive to the mass of the primary 
particle. Consequently, it is important to estimate this parameter for all events, including saturated ones, in order 
to perform reliable composition analyses. Furthermore, excluding saturated events would introduce a bias in such 
analyses based on the $\hat{\mu}(450)$ parameter, since, as shown in Fig.~\ref{SatFr}, the fraction of saturated 
events is higher for heavier primaries.

\begin{figure}[!ht]
\centering
\includegraphics[width=7.8cm]{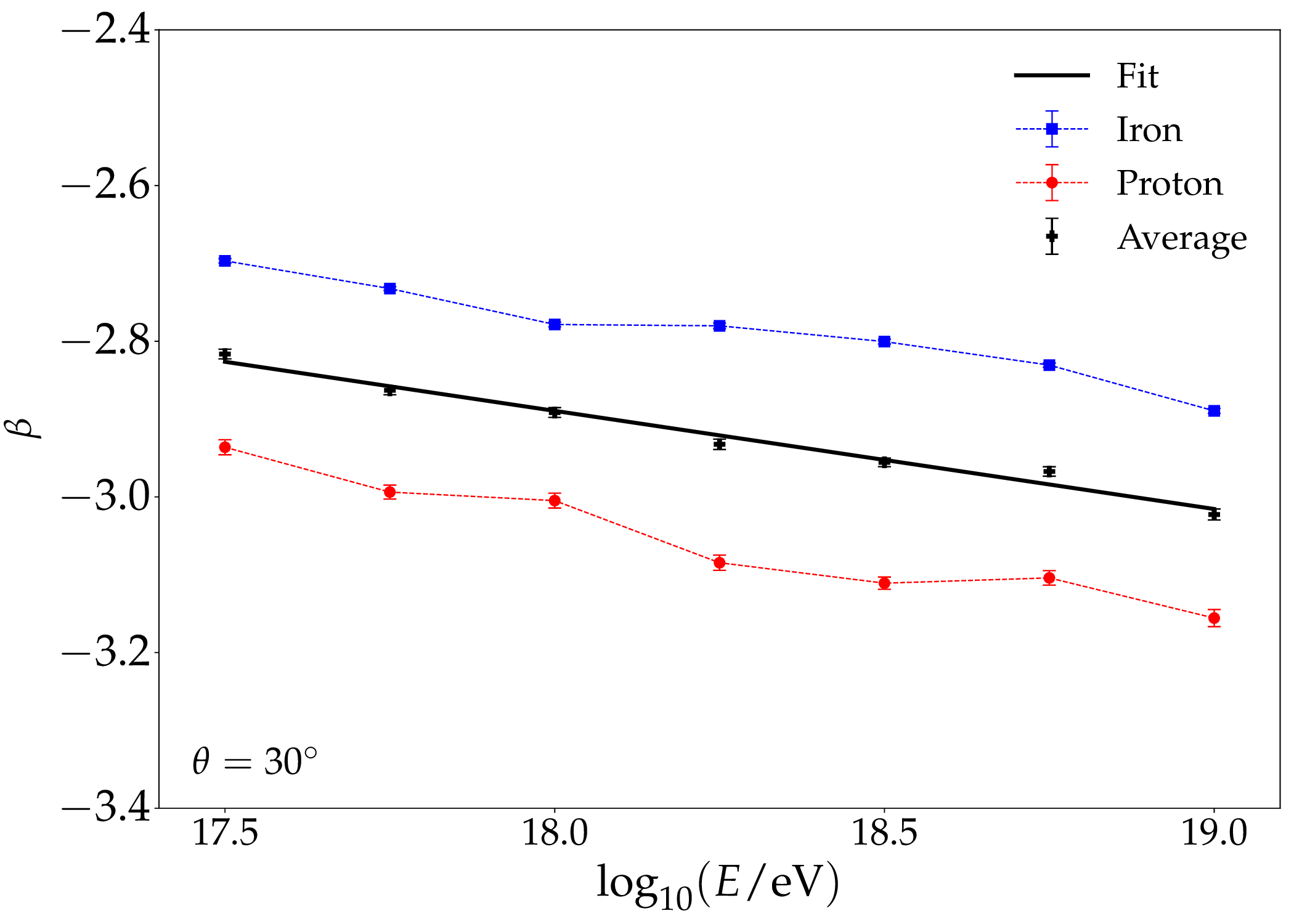}
\caption{Slope of the average MLDF for proton, iron, and a 50\% proton--iron mixture as a function of the 
logarithm of primary energy. The 50\% proton--iron mixture is fitted with a linear function. The zenith angle 
of the showers is $30^\circ$.}
\label{Beta}
\end{figure}

A potential strategy to reduce the tail of the saturated events distribution is to fix the $\beta$ parameter
during the minimization process. Figure \ref{Beta} shows the fitted values of $\beta$ as a function of the 
logarithm of the primary energy for proton, iron, and a 50\% proton–iron mixture, all at a zenith angle of 
$30^\circ$. The values of $\beta$ considered are obtained from fitting the average MLDF, as described in 
Ref.~\cite{Varada:23}. The values of $\beta$ for the 50\% proton--iron mixture are fitted with a linear 
function, which is used to determine the values of $\beta$ fixed during the reconstruction of the saturated
events. A similar fit is also done for $45^\circ$ zenith angle data and used to fix $\beta$ of the saturated
events. It is worth noting that, in practice, a parametrization of $\beta$ derived from real data can often 
be used instead, which typically yields improved results in terms of bias reduction.   

\begin{figure}[!ht]
\centering
\includegraphics[width=7.8cm]{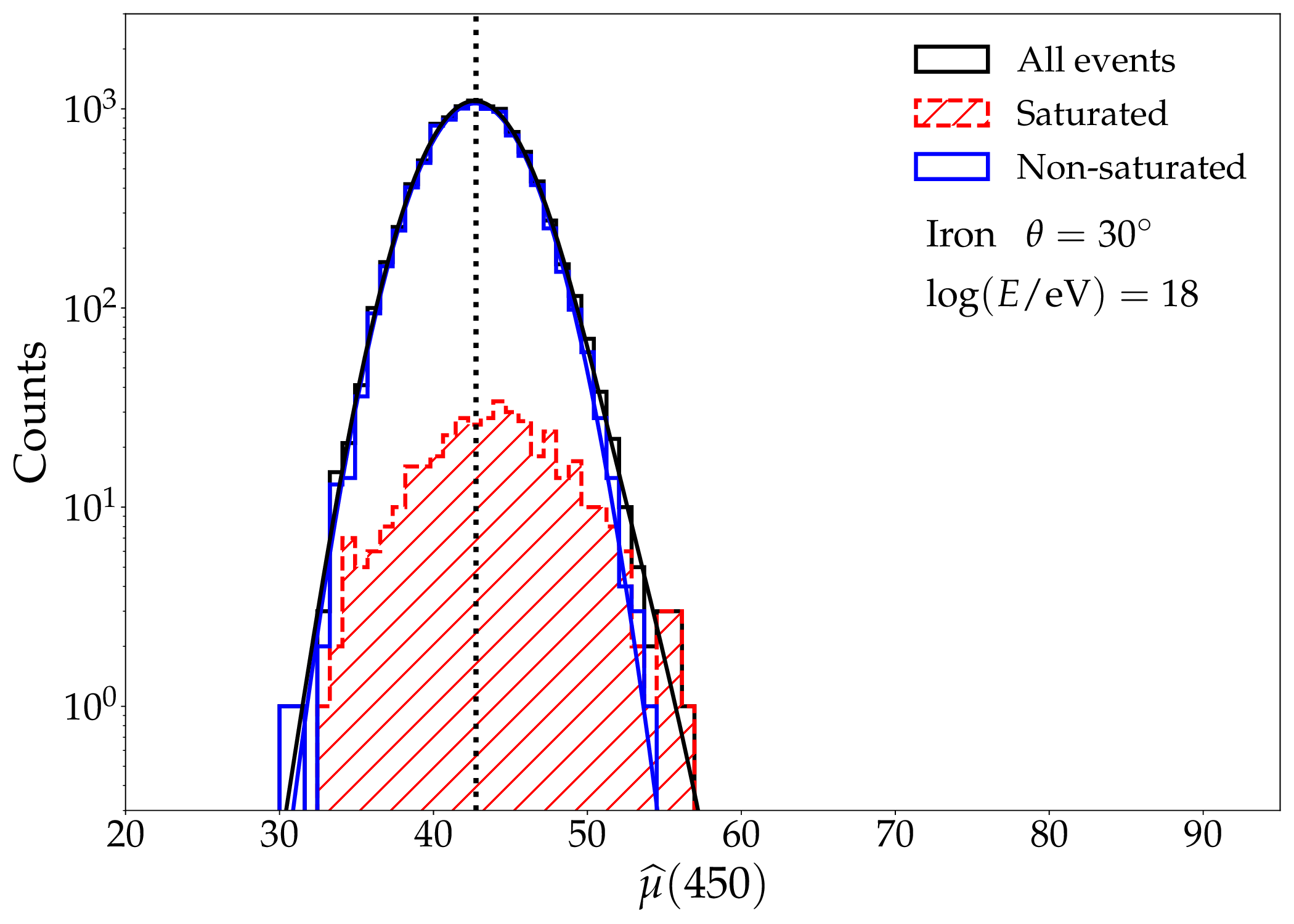}
\caption{Distributions of $\hat{\mu}(450)$ corresponding to non-saturated, saturated and all events for iron
showers of $10^{18}$ eV and 30$^\circ$ zenith angle. The blue solid line corresponds to the fit of the 
non-saturated histogram with a Gaussian function and the black solid line corresponds to the fit of the 
histogram of all events with a double-Gaussian function. The reconstruction of the saturated events is done
fixing the $\beta$ parameter in the MLDF function. The dotted vertical line represent the true value
of $\mu(450)$.}
\label{SatDistBetaFixed}
\end{figure}

Figure \ref{SatDistBetaFixed} shows the distributions of $\hat{\mu}(450)$ reconstructed using the combined method, 
this time with the parameter $\beta$ fixed for saturated events. As in Fig.~\ref{SatDist}, the distributions 
shown correspond to non-saturated, saturated, and all events from iron-induced showers with a primary energy of 
$10^{18}$ eV and a zenith angle of 30$^\circ$. The figure reveals that fixing $\beta$ significantly reduces the 
high-value tail observed in the distribution of saturated events. As a result, the single-Gaussian fit to the 
distribution of non-saturated events closely resembles the double-Gaussian fit applied to the distribution of 
all events. 

Moreover, as shown in Table~\ref{tab}, when all events are reconstructed with $\beta$ treated as a free parameter, the 
relative bias of $\hat{\mu}(450)$ becomes positive and its relative standard deviation increases. By contrast, reconstructing 
saturated events with $\beta$ fixed reduces both the relative bias and the relative standard deviation compared to the case 
where $\beta$ is left free. In this scenario, the relative standard deviation of $\hat{\mu}(450)$ is only slightly larger 
than that obtained for non-saturated events.
\begin{table*}[th!]
\centering
\caption{Relative bias (b$_i$) and relative standard deviation (RSD$_{i}$) of $\hat{\mu}(450)$ for 
non-saturated events (NS), all events reconstructed leaving $\beta$ free (All), and non-saturated events 
reconstructed leaving $\beta$ free and saturated events reconstructed fixing $\beta$ (BF). The events 
considered correspond to iron showers of $\theta=30^\circ$. }
\label{tab}
\begin{tabular}{ccccccc}
\toprule
$\log(E/\textrm{eV})$ & b$_\textrm{NS}$ [\%] & RSD$_{\textrm{NS}}$ [\%] & b$_\textrm{All}$ [\%] & RSD$_{\textrm{All}}$ [\%] & b$_\textrm{BF}$ [\%] & RSD$_{\textrm{BF}}$ [\%] \\
\midrule
18   & $\sim -0.11$ & $\sim 6.8$ & $\sim 0.17$ & $\sim 8.4$ & $\sim 0.003$ & $\sim 7.1$ \\
18.5 & $\sim -0.31$ & $\sim 4.1$ & $\sim 0.13$ & $\sim 5.8$ & $\sim 0.05$ & $\sim 4.6$ \\
\bottomrule
\end{tabular}
\end{table*}

\begin{figure}[!ht]
\centering
\includegraphics[width=7.8cm]{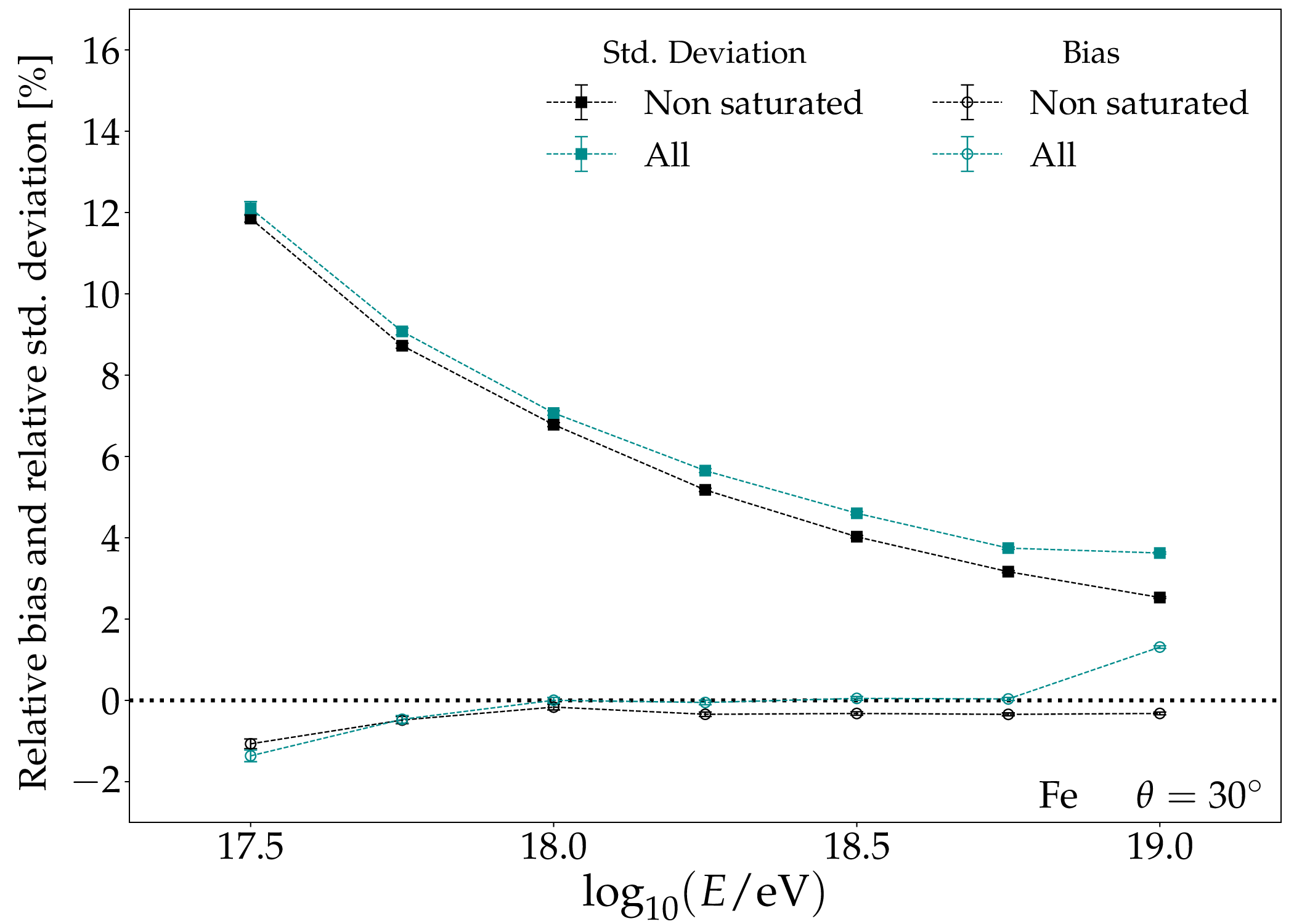}
\includegraphics[width=7.8cm]{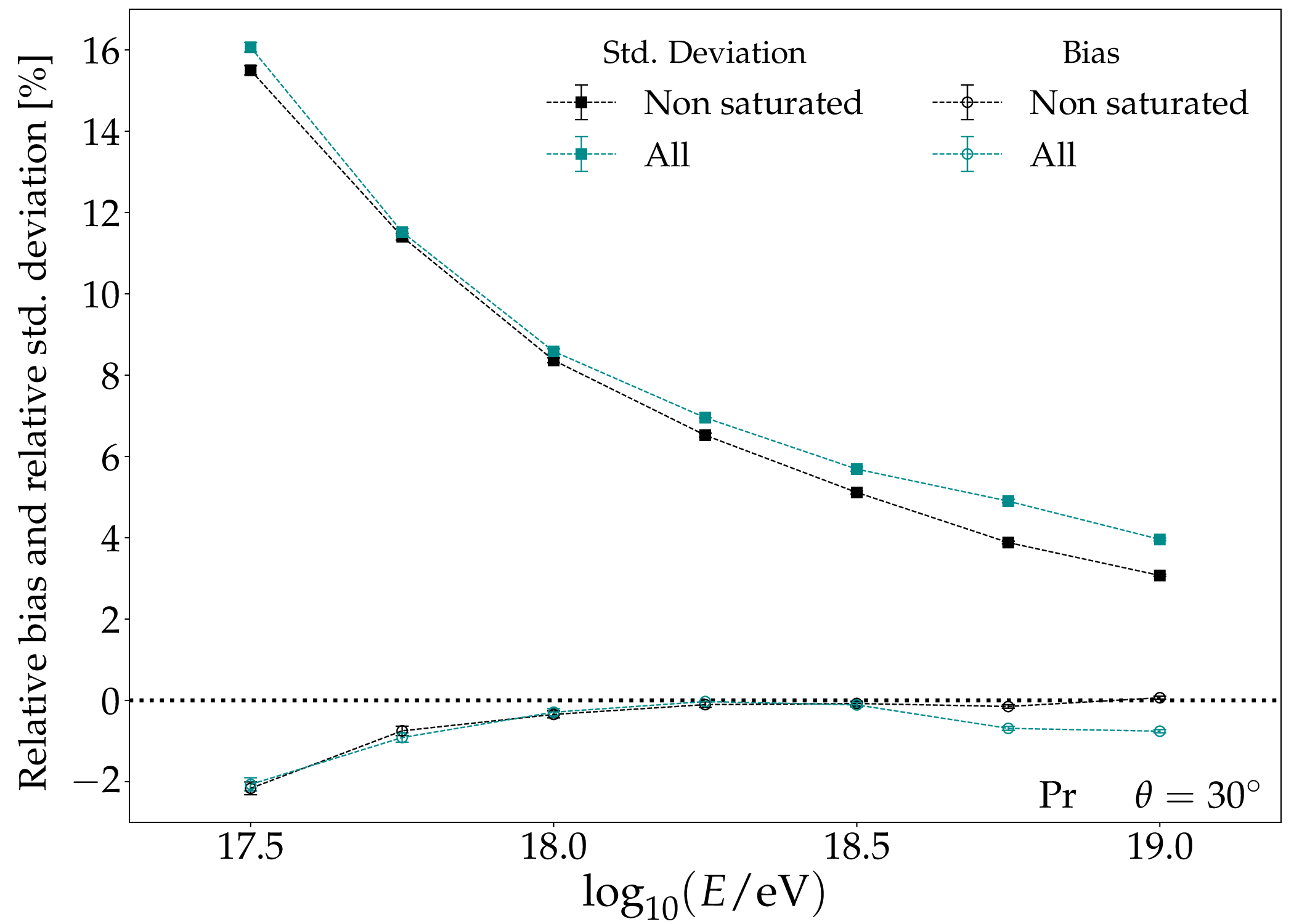}
\caption{Relative bias and relative standard deviation of $\hat{\mu}(450)$ as a function of the logarithm of the primary 
energy obtained by using the combined method. Results are shown for two event selection criteria: non-saturated events 
only and all events, where saturated events are reconstructed by fixing the slope parameter $\beta$. Top panel: Iron-induced 
air showers. Bottom panel: Proton-induced air showers. The zenith angle of the events is 30$^\circ$.}
\label{SatBetaFixed}
\end{figure}

Figure \ref{SatBetaFixed} presents the relative bias and relative standard deviation of $\hat{\mu}(450)$ as functions of 
the logarithm of the primary energy. Results are shown for two cases: one including only non-saturated events, and the 
other including all events, with saturated events reconstructed by fixing the parameter $\beta$. The top panel corresponds 
to iron-induced showers, while the bottom panel shows results for proton-induced showers, both at a zenith angle of 30$^\circ$.
The figure indicates that including all events yields a relative bias comparable to that obtained for non-saturated events. 
The relative standard deviation of $\hat{\mu}(450)$ increases slightly when saturated events are included, reflecting the 
growing fraction of saturated events with energy; however, the increase remains below $1.5\%$ across the entire energy range. 
Thus, fixing $\beta$ in the reconstruction of saturated events recovers values of the relative bias and relative standard 
deviation of $\hat{\mu}(450)$ close to those for non-saturated events. Comparable results are also obtained at a zenith 
angle of $\theta = 45^\circ$.

\section{Conclusions}
\label{sec:Conc}

In this work, we have developed a novel method to reconstruct the muon lateral distribution function using data from 
muon detectors equipped with two acquisition modes: binary and ADC. Improving reconstruction methods is crucial for composition 
analyses, as reducing uncertainties in the muon density enhances the ability to discriminate between different primary particles.
Our results show that the proposed combined method outperforms reconstructions based solely on either the binary or ADC mode. 
Furthermore, it performs better than existing hybrid approaches that select the acquisition mode based on the number 
of muons detected at each station. We also investigated the reconstruction of saturated events. We found that fixing the slope 
parameter of the muon lateral distribution function in such cases helps suppress the formation of long tails in the distribution. 
This strategy introduces only a minimal impact on the relative bias and the relative standard deviation of the reconstructed 
average muon number at the reference distance.

It is worth noting that, in this type of muon detector, the ADC is calibrated using the binary mode in the low-muon-count regime, 
where the pile-up effect is less significant and the binary mode performs better. This calibration is then extrapolated to the 
high-muon-count regime. When designing such detectors, the dynamic range of the ADC can be maximized to reduce the fraction of 
saturated events, which are more challenging to reconstruct. However, extending the calibration curve in this way also increases 
the extrapolated region, potentially introducing biases in the determination of the muon number. For this reason, a conservative 
design strategy is to set the ADC dynamic range only slightly larger than that of the binary mode.

\appendix
\section{Asymptotic expression for the Stirling number of second kind}
\label{appS2}

The computation of the Stirling number of second kind becomes computationally expensive as $n$ grows. Therefore, to speed up the 
reconstruction process an asymptotic expression is employed for large values of $n$,
which is given by \cite{S2:93}
\begin{eqnarray}
S_\textrm{app}(n,k) \! \! &=& \! \! \binom{n}{k} \, \frac{\left[ \exp(u(n/k)) - 1 \right]^k}{u(n/k)^n} \, \exp(k - n) \nonumber \\[0.1cm]
&&  (n - k)^{n - k} \, \sqrt{\frac{n - k}{n \, (u(n/k) - n/k + 1)}}, \nonumber\\
\end{eqnarray}
where the function $u$ is obtained by numerically inverting the following expression
\begin{equation}
\frac{u \exp(u)}{\exp(u)-1} =\frac{n}{k}.
\end{equation}

\begin{figure}[!ht]
\centering
\includegraphics[width=7.8cm]{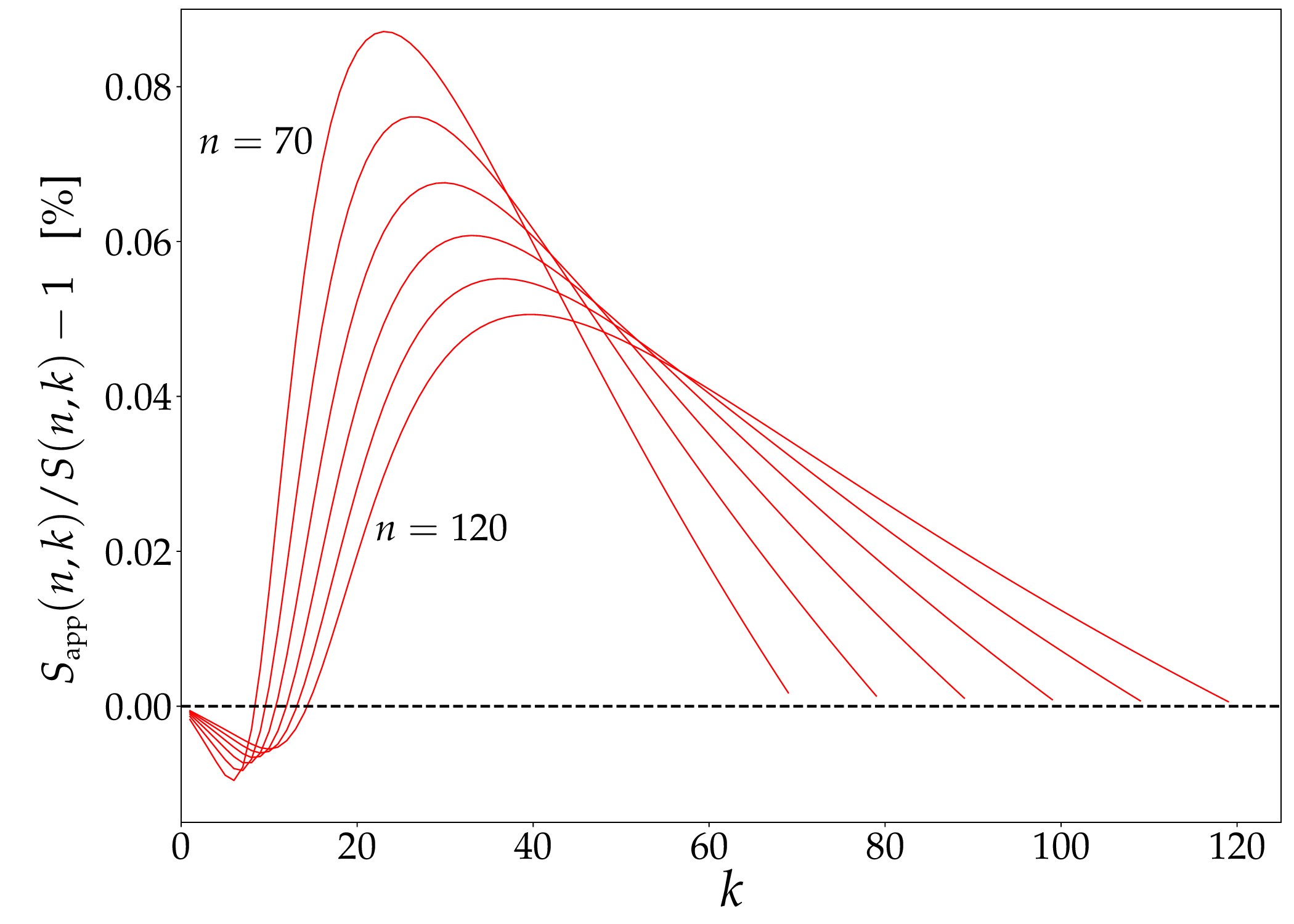}
\caption{Relative error of $S_\textrm{app}(n,k)$ as a function of for $n$ between 70 and 120 in steps of $\Delta n=10$.}
\label{S2}
\end{figure}

Figure \ref{S2} presents the relative error of $S_\textrm{app}(n,k)$ as a function of $k$, for values of $n$ ranging from 70 
to 120 in steps of $\Delta n = 10$. From the figure, it can be observed that the relative error remains below $0.09\%$ for all 
$n \geq 70$. Moreover, the relative error of $S_\textrm{app}(n,k)$ decreases consistently as $n$ increases.  

\end{document}